\documentclass[aps,pra,twocolumn,showpacs]{revtex4}
\usepackage{dcolumn}
\usepackage{multirow}
\usepackage[dvips]{graphics}
\begin{document}
[Phys. Rev. A {\bf 83}, 013616 (2011)]
\title{Three-dimensional quantum phase diagram of the exact ground states of a mixture of two species of spin-$1$ Bose gases with interspecies spin exchange}
\author{Yu Shi}
\email{yushi@fudan.edu.cn}
\affiliation{
Department of Physics, Fudan University, Shanghai 200433, China}
\author{Li Ge}
\affiliation{
Department of Physics, Fudan University, Shanghai 200433, China}

\begin{abstract}
We find nearly all the exact ground states of a mixture of two species of spin-1 atoms with both interspecies and intraspecies spin exchanges in absence of a magnetic field. The quantum phase diagram in the three-dimensional parameter space and its two-dimensional cross sections are described. The boundaries where the ground states are either continuous or discontinuous are determined, with the latter identified as where quantum phase transitions take place. The two species are always disentangled if the interspecies spin coupling is ferromagnetic or zero.  Quantum phase transitions occur when the interspecies spin coupling varies between antiferromagtic and zero or ferromagnetic  while the two intraspecies spin couplings  both remain ferromagnetic. On the other hand, by tuning the interspecies spin coupling from zero to antiferromagnetic and then back to zero, one can circumvent the quantum phase transition due to sign change of the intraspecies spin coupling of a single species, which is spin-decoupled with the other species with ferromagnetic intraspecies spin coupling. Overall speaking, interplay among interspecies and two intraspecies spin exchanges significantly enriches quantum phases of spinor atomic gases.
\end{abstract}


\maketitle

\section{Introduction}

Spinor Bose gases have been a remarkable subject since a
decade ago~\cite{ho1,law,koashi,ho2,hoyin,yang,spinor}. A mixture of two scalar Bose gases has also been studied extensively~\cite{two,myatt}. The problem of mixing two distinguishable species of
spinor Bose gases with interspecies spin exchange was first
considered in a mixture of pseudospin-$\frac{1}{2}$
gases, with particular motivation and interest in entanglement
between the two species, realizing the so-called entangled Bose-Einstein condensation~\cite{shi0,shi1,shi2,shi4}, which is a feature beyond those of scalar mixtures as well as a single species of spinor gases.  A mixture of two species of spin-$1$ atoms has also received attention more recently~\cite{luo,xu1,shi5,xu2}. Compared with a pseudospin-$\frac{1}{2}$ mixture, the spin-$1$ mixture may be more experimentally accessible and exhibits interesting properties such as the coexistence of intraspecies and interspecies singlet pairs even in a  global many-body singlet state. Previously, the exact ground states of a spin-$1$  mixture have been found in four parameter regimes, with or without a magnetic field~\cite{shi5}. This left
open to work out the ground states in other parameter regimes, especially,  an overall picture of the occupation of different ground states in the whole parameter space. However, the present problem of minimizing the energy is very complicated, with three constrained variables, namely, the total spins of the individual species and of the total mixture, as well as four independent parameters, namely, the three spin coupling strengths and the magnetic field.

In this paper, we find nearly all the possible ground states in absence of a magnetic field, by working out the complicated minimization problem (the even more difficult case with a magnetic field is discussed in a separate paper). Thus we obtain a three dimensional phase diagram, with the three spin coupling strengths as the parameters. By comparing the ground states in neighboring parameter regimes, we distinguish phase boundaries, where discontinuities of the ground states or quantum phase transitions occur, from crossover boundaries, where the ground states are continuously connected. As the spin coupling strengths, proportional to spin-exchange scattering lengths, can be tuned through Feshbach resonances,  a novel venue is opened in this system for studying quantum phase transitions and crossovers between different many-body ground states.

Consider two species $a$ and $b$ of  spin-1 atoms,  whose numbers
$N_a$ and $N_b$ are conserved respectively. The
many-body Hamiltonian has been given previously~\cite{shi5}. In absence of a magnetic field, it can be written as
\begin{equation}
{\cal H} = \frac{c^{a}}{2} \mathbf{S}_a^2 +
\frac{c^{b}}{2}\mathbf{S}_b^2 + c^{ab} \mathbf{S}_a \cdot
\mathbf{S}_b, \label{spinham}
\end{equation}
where $\mathbf{S}_{\alpha}=
\alpha^{\dagger}_{\mu}\mathbf{F}_{\mu\nu}\alpha_{\nu}$ is the total
spin operator for $\alpha$ species, $\alpha_{\mu}$ is the annihilation operator of species $\alpha$
($\alpha =a, b$),  $\mathbf{F}_{\mu\nu}$ is the
$(\mu\nu)$ element of spin-1 matrix ($\mu, \nu = -1, 0, 1$),
$c^{\alpha}$ is the intraspecies spin coupling strength of species $\alpha$, proportional to the intraspecies spin-exchange scattering length, while $c^{ab}$ is the interspecies spin coupling strength, proportional to the interspecies spin-exchange scattering length~\cite{shi5}.

$S_a$, $S_b$  and the total spin $S$ are all good quantum numbers.
Consequently, there are $2S+1$ degenerate ground states with
$S_z = -S^m, \cdots, S^m$, i.e.
$$|S_a^m,S_b^m,S^m,S_z\rangle, $$
where  ${\cal S}_a^m$, ${\cal
S}_b^m$ and ${\cal S}^m$ are, respectively, the values of $S_a$,
$S_b$ and $S$ that minimize the energy
\begin{equation}
\begin{array}{c}
\displaystyle
E=\frac{c^{a}-c^{ab}}{2 }
S_a(S_a+1)+ \frac{c^{b}-c^{ab}}{2 }S_b(S_b+1)
\\ \displaystyle
+ \frac{c^{ab}}{2
} S(S+1),
\end{array} \label{e}
\end{equation}
under the constraints
\begin{equation}
0 \leq S_{\alpha}  \leq  N_{\alpha},  \label{arange}
\end{equation}
\begin{equation}
|S_a-S_b|   \leq   S  \leq  S_a+S_b. \label{srange}
\end{equation}

In general, the problem of minimizing $E$ is very complicated, because there are three variables $S_a$, $S_b$ and $S$, with the constraints (\ref{arange}) and (\ref{srange}), as well as three parameters. Through tour de force calculations, we have done the minimization in nearly all parameter regimes, except a small regime where there are too many possibilities.  The details of solving for $S_a^m$, $S_b^m$ and $S^m$ are reported in the Appendices.

In discussing the phase diagrams, we shall focus on  the maximally polarized state $$|S_a^m,S_b^m,S^m,S^m\rangle,$$ which can be determined to be the unique ground state under an infinitesimal magnetic field.

In Sec.~\ref{gstates}, we list the ground states in different parameter regimes in a table and draw the phase diagrams on $c^a~-~c^b$ planes  for different signs of $c^{ab}$, based on calculations described in the Appendices. In Sec~\ref{threed}, we describe the  structure of the phase diagram in the three-dimensional parameter space, and its implication on effects of $c^{ab}$ on quantum phase transitions. In Sec.~\ref{comp}, we discuss composition of the ground states in terms of bosonic degrees of freedom. A summary is made in Sec~\ref{summary}.

\section{Ground states in various parameter regimes \label{gstates} }

The ground states in different parameter regimes with $c^{ab} \neq 0$ are reported in Table~\ref{table1}. Different parameter regimes are numbered according to the ordering numbers of the sections and subsections in the Appendices where the minimization of $E$ is done in these different regimes respectively.

\begin{table*}
\begin{tabular}{|l|l|l|l|}
\hline No. & \multicolumn{2}{|c|}{Parameter regimes} & {Ground states} \\
 \hline
 \multirow{2}{*}{1,A2a} &  & $c^{a} \leq 0$,
 & $|N_a,N_b,N_a+N_b, S_z \rangle$,\\
 &&$c^{b}\leq -\frac{2N_ac^{ab}}{2N_b+1}$& disentangled when $S_z=\pm(N_a+N_b)$, entangled otherwise \\ \cline{1-1} \cline{3-4}
 \multirow{2}{*}{A2b} & & $c^{a} \leq 0$, &
 $|N_a,n_1,N_a+n_1, S_z\rangle$,
 $n_1=\rm{Int}[\frac{|c^{ab}|N_a}{c^{b}}-\frac{1}{2}]$, \\
 &&$-\frac{2N_ac^{ab}}{2N_b+1} \leq c_b \leq -2N_ac^{ab}$ & disentangled when $S_z=\pm(N_a+n_1)$, entangled otherwise\\\cline{1-1} \cline{3-4}
 \multirow{2}{*}{A2c}&  & $c^{a} \leq 0$,
 & $|N_a,0,N_a, S_z \rangle = |N_a, S_z\rangle_a|0,0\rangle_b$,\\
 &&$c^{b}\geq -2N_ac^{ab} $& disentangled when $S_z=\pm N_a$, entangled otherwise \\ \cline{1-1} \cline{3-4}
  \multirow{2}{*}{1,A3a}&  & $c^{b} \leq 0$,
 & $|N_a,N_b,N_a+N_b, S_z \rangle$,\\
&&$c^{a}\leq -\frac{2N_bc^{ab}}{2N_a+1}$& disentangled when $S_z=\pm(N_a+N_b)$, entangled otherwise \\ \cline{1-1} \cline{3-4}
 \multirow{2}{*}{A3b} &  & $c^{b} \leq 0$, &
 $|n_2,N_b,n_2+N_b, S_z\rangle$,
 $n_2=\rm{Int}[\frac{|c^{ab}|N_b}{c^{a}}-\frac{1}{2}]$, \\
 & &$-\frac{2N_bc^{ab}}{2N_a+1} \leq c_a \leq -2N_bc^{ab}$ & disentangled when $S_z=\pm(n_2+N_b)$, entangled otherwise\\\cline{1-1} \cline{3-4}
  \multirow{2}{*}{A3c} &  $c^{ab} < 0$ & $c^{b} \leq 0$,
 & $|0,N_b,N_b, S_z \rangle$,\\
 &   &$c^{a}\geq -2N_bc^{ab} $& disentangled when $S_z=\pm N_b$, entangled otherwise \\ \cline{1-1}\cline{3-4}
  \multirow{2}{*}{A4}  & & $c^{a}> 0$, $0<c^{b}<-2N_ac^{ab}$,  & \\
 && $c^{a}c^{b}>(c^{ab})^2$  & \\\cline{1-1}\cline{3-3}
  \multirow{2}{*}{A5} & & $c^{b}> 0$, $0<c^{a}<-2N_bc^{ab}$,  & \\
 && $c^{a}c^{b}>(c^{ab})^2$  & \\\cline{1-1}\cline{3-3}
A6 & & $c^{a}> 0$, $c^{b}>-2N_ac^{ab}$ &  $|0,0,0,0\rangle=|0,0\rangle_a \otimes |0,0\rangle_b$, \\ \cline{1-1}\cline{3-3}
 A7 & & $c^{b}> 0$, $c^{a}>-2N_bc^{ab}$  &  disentangled \\\cline{1-1}\cline{3-3}
 \multirow{2}{*}{A8}  & & $c^{a}> 0$, $-\frac{2N_ac^{ab}}{2N_b+1} \leq c^b \leq -2N_ac^{ab}$ & \\
 &&  $c^{a}c^{b}>(c^{ab})^2$   & \\\cline{1-1}\cline{3-3}
 \multirow{2}{*}{A9} & & $c^{b}> 0$, $-\frac{2N_bc^{ab}}{2N_a+1} \leq c^a \leq -2N_b c^{ab}$, & \\
 && $c^{a}c^{b}>(c^{ab})^2$  & \\\cline{1-1}\cline{3-4}
  \multirow{2}{*}{A10}& & $0 < c^a \leq -\frac{2N_bc^{ab}}{2N_a+1} $, & $|N_a,N_b,N_a+N_b,S_z\rangle$, \\
 &&   $0 < c^b \leq -\frac{2N_ac^{ab}}{2N_b+1} $ &  disentangled when $S_z=\pm(n_2+N_b)$, entangled otherwise\\
 \hline
\multirow{2}{*}{B1} & & $c^a=c^b= c^{ab}$,
 & $|S_b^m,S_b^m,0,0\rangle$, $S_b^m=0,\cdots, N_b$ \\
 && & always entangled   \\ \cline{1-1} \cline{3-4}
 \multirow{2}{*}{B2a} & & $c^b \leq c^{ab}$,
 & $|N_a,N_b,N_a-N_b,S_z\rangle$,  \\
 &&$c^a \leq \frac{2N_bc^{ab}}{2N_a+1}$ & always entangled   \\ \cline{1-1} \cline{3-4}
 \multirow{2}{*}{B2b} & & $c^{b} \leq c^{ab}$,
 & $|n_2,N_b,n_2-N_b,S_z\rangle$,  \\
 &&
 $\frac{2N_bc^{ab}}{2N_a+1} \leq c^a \leq \frac{2N_bc^{ab}}{2N_b+1} $,  & always entangled   \\ \cline{1-1} \cline{3-4}
 \multirow{2}{*}{B2c} & &$c^{b} \leq c^{ab}$,
 & $|N_b,N_b,0,0\rangle$ \\
 &&
 $\frac{2N_bc^{ab}}{2N_b+1} \leq c^a \leq c^{ab}$,  &  entangled   \\ \cline{1-1}\cline{3-4}
 \multirow{2}{*}{B345}& $c^{ab}>0$, &$c^{a}c^{b}>{(c^{ab})^2}$
 & $|0,0,0,0\rangle=|0,0\rangle_a \otimes |0,0\rangle_b$ \\ & $N_a \geq N_b$ & $c_a>0$, $c^b>0$, $c^{ab} >0$&
 disentangled \\\cline{1-1}\cline{3-4}
 \multirow{2}{*}{B6a}&  &  $c^b \leq  0$,
 & $|N_b,N_b,0,0\rangle$ \\ && $c^{ab} < c^{a}\leq \frac{2(N_b+1)c^{ab}}{2N_b+1}$  &
 entangled  \\\cline{1-1}\cline{3-4}
 \multirow{2}{*}{B6b}&  & $c^b \leq  0$,
 & $|n_3,N_b,N_b-n_3,S_z\rangle$,
$n_3 \equiv Int[\frac{c^{ab}(N^b+1)}{c^a} -\frac{1}{2}]$ \\ && $\frac{2(N_b+1)c^{ab}}{2N_b+1} \leq c^{a} \leq 2(N_b+1) c^{ab}$ &
 entangled unless $c^{a} = 2(N_b+1) c^{ab}$, i.e. $n_3=0$  \\ \cline{1-1}\cline{3-4}\multirow{2}{*}{B6c}&  & $c^b \leq  0$,
 & $|0,N_b,N_b,S_z\rangle = |0,0\rangle_a|N_b,S_z\rangle_b$ \\ && $c^{a} \geq 2(N_b+1) c^{ab}$  &
 disentangled  \\\cline{1-1}\cline{3-4}
 \multirow{2}{*}{B7a}&  & $c^a \leq  0$,
 & $|N_a,N_b,N_a-N_b,S_z\rangle$ \\ && $c^{ab} < c^{b}\leq \frac{2(N_a+1)c^{ab}}{2N_b+1}$  &
 always entangled  \\\cline{1-1}\cline{3-4}
 \multirow{2}{*}{B7b}& & $c^a \leq  0$,
 & $|N_a,n_4,N_a-n_4,S_z\rangle$,
$n_4 \equiv \rm{Int}[\frac{(N^a+1)c^{ab}}{c^b} -\frac{1}{2}]$ \\ &&   $\frac{2(N_a+1)c^{ab}}{2N_b+1} \leq c^{b} \leq 2(N_a+1) c^{ab}$ &
 entangled unless $c^{b}= 2(N_a+1) c^{ab}$, i.e. $n_4=0$  \\ \cline{1-1}\cline{3-4}\multirow{2}{*}{B7c}&  & $c^a \leq  0$,
 & $|N_a,0,N_a,S_z\rangle=|N_a,S_z\rangle_a|0,0\rangle_b$ \\ &&  $c^{b} \geq 2(N_a+1) c^{ab}$   &
 disentangled  \\\cline{1-1}\cline{3-4}
 \multirow{2}{*}{B8}& &   $c^a > 0$, $c^b>0$,
 & $|S_a^m,S_b^m,|S_a^m-S_b^m|,S_z\rangle$, $S_a^m\neq 0$, $S_b^m\neq 0$ \\ && $c^{a}+c^b<2c^{ab}$ &
 always entangled  \\
 \hline
 \end{tabular}

\caption{Ground states of a mixture of two spin-1 atomic gases with interspecies spin coupling  $c^{ab} \neq 0$ in different parameter regimes in the absence of a magnetic field. For $c^{ab} >0$, $N_a\geq N_b$ is assumed without loss of generality. $\rm{Int}(x)$ refers to the integer closest to $x$ and in the legitimate range of the concerned spin quantum number, that is, $0 \leq n_1 \leq N_b$, $0 \leq n_2 \leq N_a$, $0 \leq n_3 \leq N_b$, $0 \leq n_4 \leq N_a$.  \label{table1}}
\end{table*}

Although the calculations are very complicated, the results are quite elegant, as displayed in the phase diagrams shown below.

\subsection{$c^{ab} < 0$}

For ferromagnetic interspecies spin coupling $c^{ab} < 0$, the maximally polarized ground state  is  in the form of \begin{equation}
|S_a^m,S_b^m,S_a^m+S_b^m,S_a^m+S_b^m\rangle =|S_a^m,S_a^m\rangle_a |S_b^m,S_b^m\rangle_b, \label{dis}
\end{equation}
which is always disentangled. However, the disentangled ground state may still be different in different parameter regimes. We can determine whether the ground states in neighboring regimes belong to a same quantum phase from whether they are continuously connected, that is, whether they both approach the ground state on the boundary between their regimes.

\begin{figure*}
\includegraphics[205,340][442,549]{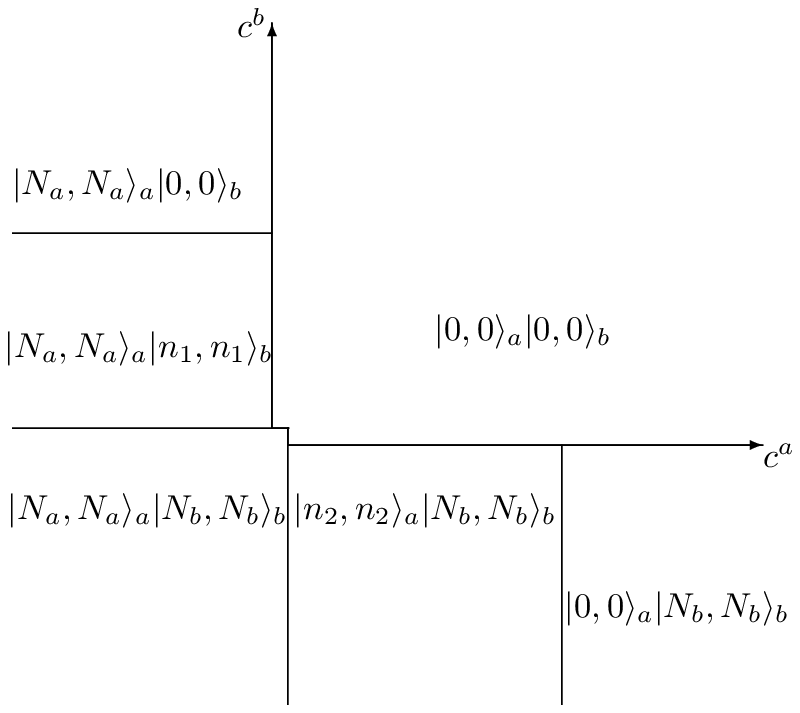}
\caption{Ground states in $c^a~-~c^b$ parameter plane for $c^{ab} <0$. $|0,0\rangle_a|0,0\rangle_b$ discontinues with all the other ground states, which belong to a same quantum phase, as they are continuous on the boundaries between them. For brevity, the coordinates of the boundaries are not labeled, as they are clearly stated in the main text and Table~\ref{table1}.  The phase boundaries belong to the phase other than  $|0,0\rangle_a|0,0\rangle_b$. \label{pd1}}
\end{figure*}

In FIG.~\ref{pd1}, we draw the two-dimensional phase diagram of the maximally polarized ground state for $c^{ab} < 0$.

The ground state is
$|0,0\rangle_a|0,0\rangle_b$ in the first quadrant $c^a >0$ and $c^b >0$ excluding the part $0 < c^a \leq \frac{-2N_bc^{ab}}{2N_a+1}$ and $0 < c^b \leq \frac{-2N_ac^{ab}}{2N_b+1}$.

In the regime $c^a \leq  \frac{-2N_bc^{ab}}{2N_a+1}$ and $c^b \leq  \frac{-2N_ac^{ab}}{2N_b+1}$, including the whole third quadrant, the ground state is  $|N_a,N_a\rangle_a|N_b,N_b\rangle_b$.

In the regime $c_a \leq 0$ and $-\frac{2N_ac^{ab}}{2N_b+1} \leq c_b \leq -2N_ac^{ab}$, the ground state is $|N_a,N_a\rangle_a|n_1,n_1\rangle_b$, which varies with $c^b$,
as $n_1 \equiv {\rm Int}[\frac{|c^{ab}|N_a}{c^{b}}-\frac{1}{2}]$ depends on $c^{ab}$ and $c^b$. It is continuously connected upwards with $|N_a,N_a\rangle_1|0,0\rangle_b$ on the boundary $c^b=-2N_ac^{ab}$, and is also continuously connected downwards with $|N_a,N_a\rangle_a|N_b,N_b\rangle_b$ on the boundary  $c^b=-\frac{2N_ac^{ab}}{2N_b+1}$.
Hence the regime of $|N_a,N_a\rangle_a|n_1,n_1\rangle_b$   acts as a crossover regime.

Likewise, in the regime $c_b \leq 0$ and $-\frac{2N_bc^{ab}}{2N_a+1} \leq c_a \leq -2N_bc^{ab}$, the ground state is $|n_2,n_2\rangle_a|N_b,N_b\rangle_b$, which varies with $c^a$,
as $n_2={\rm  Int}[\frac{|c^{ab}|N_b}{c^{a}}-\frac{1}{2}]$ depends on $c^{ab}$ and $c^a$. Throughout the paper, ${\rm  Int}(x)$ refers to the integer closest to $x$ and in the legitimate range of the concerned quantity.  This regime is continuously connected leftwards with $|N_a,N_a\rangle_a|N_b,N_b\rangle_b$ on the boundary $c^a=-\frac{2N_bc^{ab}}{2N_a+1}$, and is  continuously connected rightwards with $|0,0\rangle_a|N_b,N_b\rangle_b$  on the boundary $c^a=-2N_bc^{ab}$. Hence the regime of $|n_2,n_2\rangle_a|N_b,N_b\rangle_b$ acts as a crossover regime.

Altogether, there are only two phases for  $c^{ab} < 0$. One includes these five ground states in the second, third and fourth quadrants, including the borders $c^a=0$ and $c^b=0$,  plus  the regime $0\leq c^a\leq -\frac{2N_bc^{ab}}{2N_a+1}$ and $0\leq c^b \leq
-\frac{2N_ac^{ab}}{2N_b+1}$.
The other is just the ground state $|0,0\rangle_a|0,0\rangle_b$ in the complementary regime. A discontinuity or quantum phase transition occurs in crossing anywhere on the boundary between the two phases, which is a continuous line consisting of the intervals $c^b\geq  \frac{2N_a c^{ab}}{2N_b+1}$ while  $c^a=0$, $0< c^a \leq c^{ab}$ while
$c^b = \frac{2N_a c^{ab}}{2N_b+1}$, $0< c^b \leq c^{ab}$ while
$c^a = \frac{2N_b c^{ab}}{2N_a+1}$, and  $c^a\geq  \frac{2N_b c^{ab}}{2N_a+1}$ while  $c^b=0$.  The ground state of the total system being a product of the states of the the two species, the state of one or both species discontinue on this phase boundary. These quantum phase transitions are first order, as the total $S_z$ is discontinuous in crossing anywhere on the the phase boundary.

The ground states on the phase boundary itself need to be addressed in detail. In the present case of $c^{ab} <0$, we have carefully taken into consideration the boundaries in the calculations described in Appendix A. It turns out that the ground states on the phase boundary are always the same as the ground states on the left or downside, that is, in the phase other than $|0,0\rangle_a|0,0\rangle_b$.

\subsection{ $c^{ab} > 0$ }

\begin{figure*}
\includegraphics[127,497][438,716]{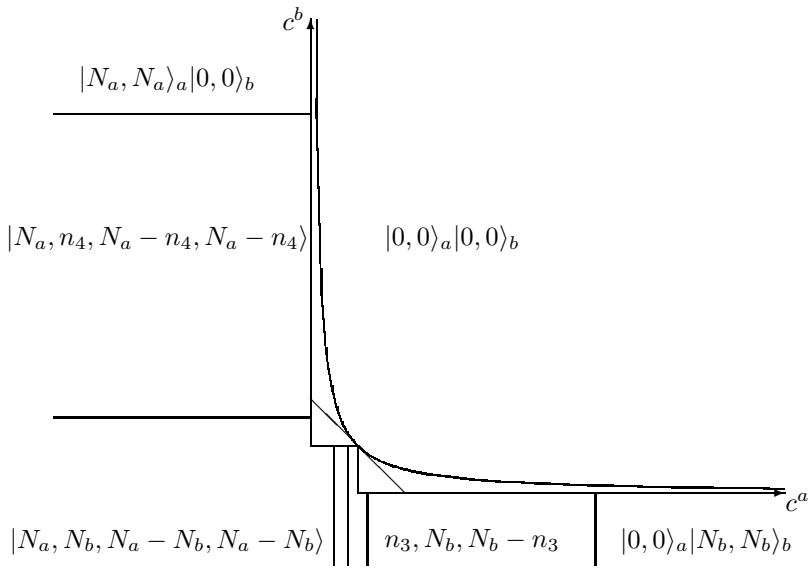}
\caption{Ground states in $c^a~-~c^b$ parameter plane for $c^{ab} > 0$.
All the ground states in the second, third and fourth quadrants plus the regime $c^a < c^{ab}$ and $ c^b < c^{ab}$ in the first quadrant  are continuous connected, forming a same quantum phase, which discontinues with $|0,0\rangle_a|0,0\rangle_b$ at $c^a = c^b = c^{ab}$. The ground states in the regime above $c^a < c^{ab}$ and $ c^b < c^{ab}$ and below $c^ac^b < (c^{ab})^2$ in the first quadrant have not determined, although it is certain that those in the subregime $c^a + c^b < 2c^{ab}$ in the first quadrant is always entangled. Limited by space, $|n_3,N_b,N_b-n_3,N_b-n_3\rangle$ is indicated by $n_3,N_b,N_b-n_3$. It is continuously connected leftwards with $|N_b,N_b,0,0\rangle$, which is continuously connected leftwards with $|n_2,N_b,n_2-N_b,n_2-N_b\rangle$, which is continuously connected leftwards with $|N_a,N_b,N_a-N_b,N_a-N_b\rangle$.  \label{pd2}}
\end{figure*}

For  antiferromagnetic interspecies spin coupling $c^{ab} > 0$, the maximally polarized ground state  is  in the form of \begin{equation}
|S_a^m,S_b^m,|S_a^m-S_b^m|,|S_a^m-S_b^m|\rangle, \label{ent}
\end{equation}
which is always entangled if
$S_a^m \neq 0$ and $ S_b^m \neq 0$. If at least one of them is $0$, then the state is disentangled as $|S_a^m,S_a^m\rangle_a|0,0\rangle_b$ or  $|0,0\rangle_a|S_b^m,S_b^m\rangle_b$.
In FIG.~\ref{pd2}, we draw the two-dimensional phase diagram of the maximally polarized ground state $|S_a^m,S_b^m,|S_a^m-S_b^m|,|S_a^m-S_b^m|\rangle$ for $c^{ab} > 0$.

Without loss of generality, we assume $N_a \geq N_b$ in the case $c^{ab} >0$.
The ground state is  $|N_a,N_b,N_a-N_b,N_a-N_b\rangle$ in the regime   $c^b \leq c^{ab}$ and $ c^a \leq \frac{2N_bc^{ab}}{2N_a+1} $  plus   $c^a \leq  0$ and $c^{ab} < c^{b}\leq \frac{2(N_a+1)c^{ab}}{2N_b+1}$. It is continuous connected upwards with $|N_a,n_4,N_a-n_4,S_z\rangle$ in the regime $c^a \leq  0$ and $\frac{2(N_a+1)c^{ab}}{2N_b+1} \leq c^{b} \leq 2(N_a+1) c^{ab}$. With
$n_4 \equiv {\rm  Int}[\frac{(N^a+1)c^{ab}}{c^b} -\frac{1}{2}]$ varying with $c^b$ from $N_b$ to $0$, $|N_a,n_4,N_a-n_4,S_z\rangle$ is  continuously connected upwards with $|N_a,N_a\rangle_a|0,0\rangle_b$ in the regime $c^a \leq  0$ and $c^{b} \geq 2(N_a+1) c^{ab}$.

To the right, $|N_a,N_b,N_a-N_b,N_a-N_b\rangle$ is continuously connected with $|n_2,N_b,n_2-N_b,n_2-N_b\rangle$ in the  regime  $c^b \leq c^{ab}$ and $\frac{2N_bc^{ab}}{2N_a+1} \leq c^a \leq \frac{2N_bc^{ab}}{2N_b+1} $.  As $n_2 \equiv {\rm  Int}[\frac{|c^{ab}|N_b}{c^{a}}-\frac{1}{2}]$ varies from $N_a$ to $N_b$, depending on $c^a$, $|n_2,N_b,n_2-N_b,n_2-N_b\rangle$  is also continuously connected rightwards with $|N_b,N_b,0,0\rangle$ in the regime
$c^{b} \leq c^{ab}$ and $\frac{2N_bc^{ab}}{2N_b+1} \leq c^a \leq c^{ab}$ (excluding $c^a=c^b=c^{ab}$) plus
$c^b \leq  0$ and $c^{ab} < c^{a} \leq \frac{2(N_b+1)c^{ab}}{2N_b+1}$.
The latter part is further continuously connected rightwards with $|n_3,N_b,N_b-n_3,N_b-n_3\rangle$ in
the regime $c^b \leq  0$ and  $\frac{2(N_b+1)c^{ab}}{2N_b+1} \leq c^{a} \leq 2(N_b+1) c^{ab}$. As $n_3 \equiv {\rm  Int}[\frac{c^{ab}(N^b+1)}{c^a} -\frac{1}{2}]$ varies from $N_b$ to $0$,
$|n_3,N_b,N_b-n_3,N_b-n_3\rangle$  is also  continuously connected rightwards with  $|0,0\rangle_a|N_b,N_b\rangle_b$ in the regime $c^b \leq  0$ and $c^{a} \geq 2(N_b+1) c^{ab}$. Note that the apparent asymmetry  between $a$ and $b$ is due to $N_a \geq N_b$ by definition. If $N_a=N_b$, then the phase diagram structure is indeed symmetric between $a$ and $b$, as the regime of $|n_2,N_b,n_2-N_b,n_2-N_b\rangle$ disappears, while $|N_b,N_b,0,0\rangle=|N_a,N_b,N_a-N_b,N_a-N_b\rangle$ and thus their  regimes merge.

Therefore, for $c^{ab} >0$, on   $c^a~-~c^b$ parameter plane, the second, third and fourth quadrants, plus the regime $c^a \leq c^{ab}$ and $c^b \leq c^{ab}$ (excluding $c^a=c^b=c^{ab}$)  in the first quadrant,  form a single quantum phase. Within this phase, an
entangled state $|N_a,N_b,N_a-N_b,N_a-N_b\rangle$ on the lower left part  crossovers upwards finally to the disentangled states  $|N_a,N_a\rangle_a|0,0\rangle_b$, and rightwards finally to the disentangled state $|0,0\rangle_a|N_b,N_b\rangle_b$.

This regime of $|N_b,N_b,0,0\rangle$  neighbors the regime $c^a c^b > (c^{ab})^2$, $c^a >0$ and $c^b >0$,  where the ground state is $|0,0\rangle_a|0,0\rangle_b$, on $c^a = c^b = c^{ab}$, where there are degenerate ground states $|S_b^m,S_b^m,0,0\rangle$ with $S_b^m=0,\cdots,N_b$, two of which are the ground states in the two regimes respectively. Starting from one of these two regimes and  varying the parameters towards the boundary $c^a = c^b = c^{ab}$,  the ground state remains as the starting one on the boundary $c^a = c^b = c^{ab}$, and then jumps to the ground state in the other regime upon entering it. As $S=S_z=0$ in both $|N_b,N_b,0,0\rangle$ and $|0,0\rangle_a|0,0\rangle_b$, the quantum phase transition is a continuous phase transition.

Unfortunately, we did not work out the detailed ground states in the part of the first quadrant with $c^a c^b \leq (c^{ab})^2$ while $c^a > c^{ab}$ and $c^b > c^{ab}$,
as the situation is too complicated. However, we have shown in the Appendix~\ref{apb8} that in the sub-regime with $c^a>0$, $c^b >0$ and $c^a+c^b<2c^{ab}$, the ground state is always the entangled state of the form $|S_a^m,S_b^m,|S_a^m-S_b^m|,|S_a^m-S_b^m|\rangle$, where $S_a^m \neq 0$ and $S_b^m \neq 0$. This is in consistency with the states $|N_a,N_b,N_a-N_b,N_a-N_b\rangle$ and  $|n_2,N_b,N_b-n_2,N_b-n_2\rangle$ whose regimes  overlap with the present one.

$c^ac^b=(c^{ab})^2$ is an equilateral hyperbola with $c^a$ and $c^b$ coordinate axes as the asymptotes, and the length of the real axis $\sqrt{2}c^{ab}$.

\subsection{ $c^{ab} =0$}

If $c^{ab}=0$, then ${\cal H}
= {\cal H}_a + {\cal H}_b$, where ${\cal H}_{\alpha}$ is the
$\alpha$-species part of the Hamiltonian ($\alpha =a, b$). Consequently the ground
state is the direct product $|S_a^m,S_{az}\rangle|S_b^m,S_{bz}\rangle$ of the respective degenerate ground
states of ${\cal H}_{a}$ and ${\cal H}_b$, with $S_a^m$ and $S_b^m$ determined by minimizing them independently.
The maximally polarized ground state is $|S_a^m,S_a^m\rangle|S_b^m,S_{b}^m\rangle$ is of the same form (\ref{dis}) as in the case of  $c^{ab} < 0$.

\begin{figure*}
\includegraphics[224,508][438,714]{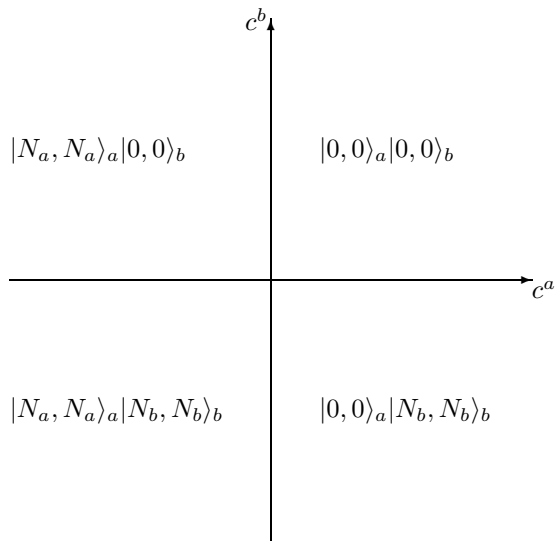}
\caption{\label{pd3} Ground states in $c^a-c^b$ parameter plane for $c^{ab} = 0$, which are direct products of the ground states of the two species. If $c^{\alpha}=0$, then the ground state of species $\alpha$ is degenerate with any legitimate value of $S_{\alpha}$, ($\alpha= a, b$).   }
\end{figure*}

As depicted in FIG.~\ref{pd3}, the phase diagram of the maximally polarized ground states for $c^{ab} =0$
simply consists of four quadrants, in each of which the ground state is a product state, that is, $|0,0\rangle_a|0,0\rangle_b$ for $c^a >0$ and $c^b >0$, $|N_a,N_a\rangle_a|0,0\rangle_b$ for $c^a <0$ and $c^b >0$, $|0,0\rangle_a|N_b,N_b\rangle_b$ for $c^a >0$ and $c^b <0$, and $|N_a,N_a\rangle_a|N_b,N_b\rangle_b$ for $c^a <0$ and $c^b <0$. When $c^{\alpha}=0$, the ground state of species $\alpha$ is a degenerate one $|S_{\alpha},S_{\alpha}\rangle$, with any legitimate value of $S_{\alpha}$, while the ground state of the other species is still determined by the sign of its intraspecies spin coupling. With $c^{ab}=0$, the origin $c^a=c^b=0$ is a quadra-critical point, where the ground state is any state of the form of $|S_a,S_a\rangle|S_b,S_b\rangle$, degenerate in both
$S_a$ and $S_b$, or any of the superposition of these states.  With the discontinuities of ground state and of the total $S_z$, a first order quantum phase transition occurs in crossing  anyway on the boundaries. If the ground state is not limited to be maximally polarized, then it can be $|S_a,S_{az}\rangle|S_b,S_{bz}\rangle$ with arbitrary legitimate values of  $S_a$, $S_{az}$, $S_b$ and $S_{bz}$ or any of their superposition.

\section{Three dimensional phase diagram \label{threed}}

\subsection{Structure}

With the two-dimensional $c^a~-~c^b$ phase diagrams given above for all possible values of $c^{ab}$, we actually have obtained the three dimensional phase diagram with $c^a$, $c^b$ and $c^{ab}$ being the three coordinates. It is very difficult to draw such a three-dimensional phase diagram, so we only make some descriptions in the following.

In the three dimensional parameter space, a boundary  $c^{\beta} = \gamma c^{ab}$   represents a plane intersecting the plane $c^{ab}=0$ on $c^{\alpha}$ axis  and with an angle $\arctan (1/\gamma)$. Here $\alpha \neq \beta$ represent the two species,   $\gamma$ represents a coefficient,

In the half space of $c^{ab}<0$, the phase boundary separating $|0,0\rangle_a|0,0\rangle_b$ from the other phase is given by the planes $c_b= -\frac{2N_ac^{ab}}{2N_b+1}$  and $c_a= -\frac{2N_bc^{ab}}{2N_a+1}$ in the fifth octant ($c^a >0$, $c^b >0$, $c^{ab}<0$), together with the plane $c^a=0$ for $c_b > -\frac{2N_ac^{ab}}{2N_b+1}$ and the plane $c^b =0$ for $c_a >  -\frac{2N_bc^{ab}}{2N_a+1}$.  Within the other phase, the crossover boundaries are planes $c_b = -\frac{2N_ac^{ab}}{2N_b+1}$ and $c_b=-2N_ac^{ab}$ in the sixth octant ($c_b >0$, $c_a<0$,  $c^{ab} <0$), and planes  $c_a = -\frac{2N_bc^{ab}}{2N_a+1}$ and $c_a= - 2N_b c^{ab}$ in the eighth octant ($c_a<0$, $c_b <0$, $c^{ab} <0$).

Now consider the  half space  $c^{ab}>0$.  The boundary surface  of $|0,0\rangle_a|0,0\rangle_b$ is a quarter of the hyperbolic paraboloid $c^a c^b = (c^{ab})^2$, with all the three coordinates positive, i.e. in the first octant. Its intersection figure with a fixed value of $c^{ab}$ gives the hyperbola boundary in the two-dimensional diagram as discussed in the last section. The intersection figure with a fixed value of $c^b$ is half of a parabola with the vertex at $c^a=c^{ab}=0$, i.e. on the $c^b$ axis, and focus distance $c^b/4$. The intersection figure with a fixed value of $c^a$ is half of a parabola with the vertex at $c^b=c^{ab}=0$. i.e. on the $c^a$ axis, and focus distance $c^a/4$.

For  $c^{ab}>0$, the crossover boundaries $c^a = \frac{2N_bc^{ab}}{2N_a+1}$, $c^a = \frac{2N_bc^{ab}}{2N_b+1}$ and $c^a = c^{ab}$ under the constraint $c^{b} \leq c^{ab}$, as well as
$c^a = \frac{2(N_b+1)c^{ab}}{2N_b+1}$ and  $c^{a} = 2(N_b+1) c^{ab}$ under the constraint  $c^b \leq  0$  are all planes starting from, but excluding,  $c^b$ axis and extending to infinity.
Likewise, the crossover boundaries $ c^b = c^{ab}$, $ c^{b} = \frac{2(N_a+1)c^{ab}}{2N_b+1}$, $c^{b} = 2(N_a+1) c^{ab}$ under the constraint $c^a \leq  0$ are all planes starting from, but excluding, the $c^a$ axis and extending to infinity.

$c^a=c^b= c^{ab} > 0$, where the two phases neighbor and  there are degenerate ground states $|S_b^m,S_b^m,0,0\rangle$,  now represents a straight line starting from the origin and extending in the first octant. $c^{a}+c^b<2c^{ab}$,  $c^a >0$ and $c^b >0$ now represent the part of the first octant that is surrounded by plane $c^{a}+c^b=2c^{ab}$, which starts from the origin, together the planes $c^a=0$ and $c^b=0$.

\begin{figure*}
\includegraphics[212,507][666,715]{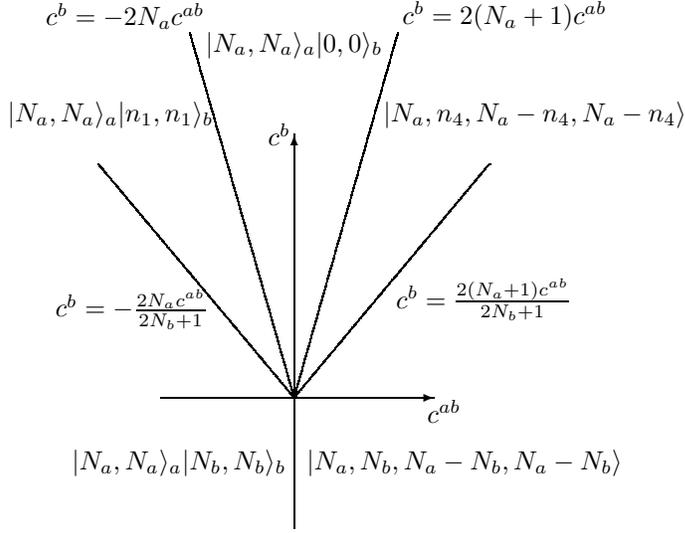}
\caption{\label{cs1} $c^b-c^{ab}$ phase diagram for a given $c^a<0$. The ground state is continuous on the four crossover boundaries represented as the four oblique lines (excluding $c^b=c^{ab}=0$), but discontinues on the negative half of $c^b$ axis.}
\end{figure*}

In FIG.~\ref{cs1}, we draw the two-dimensional $c^b-c^{ab}$ cross section for a given $c^a < 0$. The intersection figures with the continuous connecting boundaries are the four oblique lines starting from, but excluding, $c^b=c^{ab}=0$ (i.e. $c^a$ axis). The ground states on the two sides of each of these four  boundaries are continuously connected. But $|N_a,N_a
\rangle_a|N_b,N_b\rangle_b$ and $|N_a,N_b,N_a-N_b,N_a-N_b
\rangle$ discontinue on the negative half of $c^b$ axis  ($c^b < 0$, $c^{ab}=0$).  At $c^b=c^{ab}=0$, the maximally polarized ground states are degenerate, being $|N_a,N_a
\rangle_a|S_b,S_b\rangle_b$, where $S_b$ is any legitimate spin quantum number of $b$ species.

\begin{figure*}
\includegraphics[216,507][508,715]{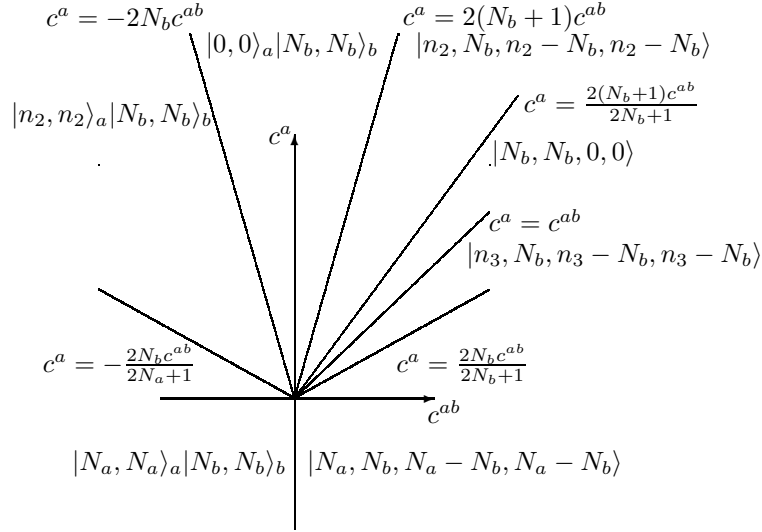}
\caption{\label{cs2} $c^a-c^{ab}$  phase diagram for a given $c^b<0$. The ground state is continuous on the four crossover boundaries represented by the six oblique lines (excluding $c^a=c^{ab}=0$), but discontinues on the negative half of $c^b$ axis. }
\end{figure*}

Similarly, in FIG.~\ref{cs2}, we draw the two-dimensional $c^a-c^{ab}$ cross section  for a given $c^b < 0$. The intersection figures with the crossover boundaries are given as six oblique lines (excluding the origin $c^b=c^{ab}=0$). The ground stats on the two sides of each of these six boundaries are continuously connected. But $|N_a,N_a
\rangle_a|N_b,N_b\rangle_b$ and $|N_a,N_b,N_a-N_b,N_a-N_b
\rangle$ discontinue on the negative half of $c^a$ axis  ($c^a < 0$, $c^{ab}=0$).  At $c^a=c^{ab}=0$, the maximally polarized ground states are degenerate, being $|S_a,S_a\rangle_a |N_b,N_b
\rangle_a$, where $S_a$ is any legitimate spin quantum number  of $a$ species.

\subsection{Convergence of the boundaries as $c^{ab} \rightarrow 0$}

In the two-dimensional $c^a-c^b$ phase diagram with each value of $c^{ab}$, the distance from $c^a=c^b=0$ to each boundary between two ground states  are proportional to $|c^{ab}|$.

In the three-dimensional parameter space, as $c^{ab} \rightarrow 0 $, all   boundaries evolve to $c^a~-~c^b$ axes. All boundaries of the form  $c^{\alpha} = \gamma c^{ab}$ evolve toward
$c^{\alpha}=c^{ab}=0$, i.e. $c^\beta$ axis ($\beta \neq \alpha$).  $c^ac^b =(c^{ab})^2$ evolves towards $c^a$ and $c^b$ axes. The lines $c^a=c^b=c^{ab}$ and  $c^a+c^b=2c^{ab}$ in the first octant both evolve towards the origin $c^a=c^b=c^{ab}=0$.

Therefore, as $c^{ab} \rightarrow 0 $, the boundaries between different ground states converge to those of $c^{ab}=0$.

However, as $c^{ab} \rightarrow 0 $, the crossover regimes in $c^a-c^b$ phase diagrams diminish, and the same phase in the second, third and fourth quadrants tend to be separated to three phases. Indeed, as we know,  there is no crossover regime in $c^a-c^b$ phase diagram for $c^{ab} = 0$. In a $c^a-c^{ab}$ cross-section for $c^a <0$ (FIG.~\ref{cs1}) and in a $c^b-c^{ab}$ cross-section for $c^b <0$ (FIG.~\ref{cs2}), the boundary lines of continuous connection  converge to the respective origins.

\subsection{Quantum phase transitions caused by varying $c^{ab}$}

The ground states on $c^{ab}=0$ plane is always continuous with those for $c^{ab}<0$, because  when $c^{ab}=-\epsilon$, where $\epsilon$ represents a positive infinitesimal quantity, the ground states in each regime of $c^a$ and $c^b$ is the same as $c^{ab}=0$. Moreover,
in first, second and fourth quadrants, the ground states on $c^{ab}=0$ plane is also continuous with those for $c^{ab}>0$.

The only exception is that the ground state on the third quadrant of $c^{ab}=0$ plane is discontinuous with that for $c^{ab}>0$,  although the boundary of the $c^a-c^b$ regime is continuous from $c^{ab}>0$ to $c^{ab}=0$. The ground state is  $|N_a,N_a,N_a-N_b,N_a-N_b\rangle$ if $c_{ab} > 0$, $c^a \leq 0$ and $c^b \leq 0$, but is  $|N_a,N_a\rangle_a|N_b,N_b\rangle_b$ if $c_{ab} \leq 0$,  $c^a < 0$ and $c^b < 0$  or degenerate product states if $c_{ab}$ and one or both of $c^a$ and $c^b$ are zero. Hence there is always a discontinuity when $c^{ab}$ changes its sign from positive to zero or negative, while $c^a$ and $c^b$ are negative or zero.

Therefore there is a quantum phase transition in crossing anywhere on the quarter plane $c^{ab}=0$, $c^a \leq 0$, $c_b \leq 0$ (strictly speaking, between $c^{ab} >0 $ and $c^{ab} \leq 0 $), depicted as the negative half of $c^b$ axis in FIG.~\ref{cs1} and the negative half of $c^b$ axis in FIG.~\ref{cs2}.
This is the only regime where quantum phase transitions are due to variation of interspecies spin coupling strength. This quantum phase transition is first order, as there is a discontinuity of $S_z$, which is  $N_a+N_b$ for $c^{ab}\leq 0$, but is $|N_a-N_b|$ for $c^{ab} >0$.

However, in the three-dimensional parameter space, these two ground states can be continuously related  through the crossover regimes with $c^a <0$ and $c^b>0$, as indicated in FIG.~\ref{cs1}, or  through the crossover regimes with  $c^b <0$ and $c^a>0$,  as indicated in FIG.~\ref{cs2}. In other words, the quantum phase transition can be circumvented if the sign of $c^{ab}$ is changed while  one of $c^a$ and $c^b$ is changed from negative to positive and then back.

\subsection{Avoiding single-species quantum phase transition by tuning $c^{ab}$}

In the three-dimensional parameter space, the part of the first octant  with $0\leq  c^a \leq c^{ab}$ and $0\leq c^b \leq
c^{ab}$ (excluding $c^a=c^b=c^{ab}$), the part of the fifth octant with $0\leq c^a\leq -\frac{2N_bc^{ab}}{2N_a+1}$ and $0\leq c^b \leq
-\frac{2N_ac^{ab}}{2N_b+1}$, and all the other six octants correspond to a same quantum phase. The reason is the following.

The ground state $|N_a,N_a\rangle_a|0,0\rangle_b$ occupies
the regime $c^{a} \leq 0$ and  $c^{b}\geq -2N_ac^{ab}$ in  $c^{ab} <0$ half space, the whole second quadrant of the $c^{ab}=0$ plane,  as well as the regime
$c^a \leq  0$ and   $c^{b} \geq 2(N_a+1) c^{ab}$ in  $c^{ab} >0$ half space. In other words, $|N_a,N_a\rangle_a|0,0\rangle_b$ occupies the regime  $- c^{b}/(2N_a)\leq c^{ab} \leq c^b/[2(N_a+1)]$ in the $c^a <0$ and $c^b>0$ subspace, fully including the second quadrant of the $c^{ab}=0$ plane (see FIG.~\ref{cs1}).
Similarly, the ground state $|0,0\rangle_a|N_b,N_b\rangle_b$ occupies the regime  $- c^{a}/(2N_b)\leq c^{ab} \leq c^a/[2(N_b+1)]$ in the $c^a >0$ and $c^b <0$ half space, fully including the fourth quadrant of the $c^{ab}=0$ plane  (see FIG.~\ref{cs2}).

Consequently some quantum phase transitions that are inevitable when $c^{ab}=0$ can be avoided by introducing negative $c^{ab}$.
With $c^{ab} = 0$, there is a first order quantum phase transition when $c^b$ changes its sign while $c^a <0$, from $|N_a,N_a\rangle_a|0,0\rangle_b$ to $|N_a,N_a\rangle_a|N_b,N_b\rangle_b$,  as indicated in FIG.~\ref{pd3} (see also $c^b$ axis in FIG.~\ref{cs1}).
These two ground states can be continuously related through the crossover regime with $c^a<0$ and $c^{ab} <0$ in the three dimensional parameter space (see FIG.~\ref{cs1}).
Similarly, with $c^{ab} = 0$, there is a quantum phase transition when $c^a$ changes its sign while $c^b <0$, from $|0,0\rangle_b|N_b,N_b\rangle_a$ to $|N_a,N_a\rangle_a|N_b,N_b\rangle_b$,  as indicated in FIG.~\ref{pd3} (see also $c^a$ axis in FIG.~\ref{cs2}).
These two ground states can be continuously related through the crossover regime with $c^b<0$ and $c^{ab} <0$ in the three dimensional parameter space (see FIG.~\ref{cs2}).

This implies that the quantum phase transition of a first species of spin-1 atoms, in varying the intraspecies spin-exchange scattering length between  negative and positive,  can be circumvented by mixing it with a second species of spin-1 atoms with negative intraspecies spin-exchange scattering length, and then tuning the interspecies spin exchange scattering length from zero to negative and then back, in the same time as the intraspecies spin-exchange scattering length of the first species varies.

\section{Composite structures in Bosonic degrees of freedom \label{comp}}

Any state $|S_a,S_b,S,S_z\rangle$ of the mixture can be written in terms of bosonic operators, as
\begin{equation}
|S_a,S_b,S,S_z\rangle = \sum_{S_{bz}} g(S_{bz}) |S_a,S_z-S_{bz}\rangle_a
|S_b,S_{bz}\rangle_b,
\end{equation}
where\begin{equation}|S_{\alpha},S_{\alpha z}\rangle_{\alpha}=[r(S_{\alpha})\cdots r(S_{\alpha
z}+1)]^{-1} (S_{\alpha_-})^{S_{\alpha}-S_{\alpha
z}}|S_{\alpha},S_{\alpha}\rangle_{\alpha},
\end{equation} with
\begin{equation}S_{\alpha_-} = \sqrt{2}(\alpha_1^{\dagger}\alpha_0 +
\alpha_0^{\dagger}\alpha_{-1}),
\end{equation} \begin{equation}
r(m_{\alpha}) \equiv \sqrt{(S_{\alpha}+m)(S_{\alpha}-m_{\alpha} +1)},
\end{equation}
and \begin{equation}
|S_{\alpha},S_{\alpha}\rangle_{\alpha}=[M(S_{\alpha})]^{-1/2} {\alpha_1^{\dagger}}^{S_{\alpha}} {\Theta_{\alpha}^{\dagger}}^{(N_{\alpha}-S_{\alpha})/2}|0\rangle
\end{equation} subject to the condition that $N_{\alpha}-S_{\alpha}$ is even,  \begin{equation}\Theta_{\alpha}^{\dagger} \equiv \alpha_1^{\dagger}\alpha_{-1}^{\dagger}
-{\alpha_0^{\dagger}}^2
\end{equation} is the creation operator of an intraspecies singlet pair of species $\alpha$,  $M(S_{\alpha})$ is the normalization constant, $g(S_{bz})$ is the standard Clebsch-Gordan coefficient~\cite{cg}
\begin{widetext}
\begin{equation}
\begin{array}{l}
\displaystyle
g(S_{bz})= \langle S_a,S_z-S_{bz}; S_b,S_{bz}|S_a,S_b,S,S_{z}\rangle
= \displaystyle \left[ \frac{(2S+1)(S_a+S_b-S)!(S_a-S_b+S)!(S_b-S_a+S)!}{(S_a+S_b+S+1)!} \right]^{\frac{1}{2}} \\ \displaystyle
\times [(S_a+S_z-S_{bz})!(S_a-S_{z}+S_{bz})! (S_b+S_{bz})! (S_b-S_{bz})! (S+S_z)! (S-S_z)! ]^{\frac{1}{2}} \\
\times \sum_k (-1)^{k}[k!(S_a+S_b-S-k)!(S_a-S_z+S_{bz}-k)!
(S_b+S_{bz}-k)!(S-S_b+S_z-S_{bz}+k)!(S-S_a-S_{bz}+k)!]^{-1},
\end{array}
\end{equation}
where $k$ is an integer such that the arguments in the factorials are non-negative. The summation over $k$ reduces to only one term  in the maximally polarized ground states in each parameter regime, and in the ground states for $c^{ab} > 0$.

If $S_z=S$, we have
\begin{equation}
\begin{array}{l}
\displaystyle
g(S_{bz})=\langle S_a,S-S_{bz}; S_b,S_{bz}|S_a,S_b,S,S\rangle\\
\displaystyle = (-1)^{S_{bz}} \left[ \frac{(2S+1)!(S_a+S_b-S)!(S_a-S_{bz}+S)!(S_b+S_{bz})! }{(S_a+S_b+S+1)!(S_a-S_b+S)!
(S_a-S+S_{bz})!(S_b-S_{bz})!} \right]^{\frac{1}{2}},
\end{array}
\end{equation}
which gives $g_{IV}(S_{bz})$ in Ref.~\cite{shi4} by substituting $S_a$, $S_b$ and $S$ as $N_a$, $N_b$ and $n$, respectively.

If $S=S^a-S^b \geq 0$, we have
\begin{equation}
\begin{array}{l}
\displaystyle
g(S_{bz})=
\langle S_a,S_z-S_{bz}; S_b,S_{bz}|S_a,S_b,S_a-S_b,S_z\rangle\\
\displaystyle = (-1)^{S_{bz}} \left[ \frac{(2S_a-2S_b+1)!(2S_b)!(S_a+S_z-S_{bz})!(S_a-S_{z}+S_{bz})!}
{(2S_a+1)!(S_b+S_{bz})!(S_b-S_{bz})!
(S_a-S_b+S_{z})!(S_a-S_b-S_{z})!} \right]^{\frac{1}{2}}.
\end{array}
\end{equation}
In particular, for $S_z=S=S_a-S_b$,
\begin{equation}
\begin{array}{l}
\displaystyle
g(S_{bz})=\langle S_a,S_a-S_b-S_{bz}; S_b,S_{bz}|S_a,S_b,S_a-S_b,S_a-S_b\rangle
= (-1)^{S_{bz}} \left[ \frac{(2S_a-S_b+1)(2S_b)!(2S_a-S_b-S_{bz})!}
{(2S_a+1)!(S_b-S_{bz})!} \right]^{\frac{1}{2}}.
\end{array}
\end{equation}

If $S=S_b-S_a\geq 0$, we have
\begin{equation}
\begin{array}{l}
\displaystyle
g(S_{bz})=\langle S_a,S_z-S_{bz}; S_b,S_{bz}|S_a,S_b,S_b-S_a,S_z\rangle\\ \displaystyle
= (-1)^{S_{bz}} \left[ \frac{(2S_b-2S_a+1)!(2S_a)!(S_b+S_{bz})!(S_b-S_{bz})!}
{(2S_b+1)!(S_a+S_z-S_{bz})!(S_a-S_{z}+S_{bz})!
(S_b-S_a+S_{z})!(S_b-S_a-S_{z})!} \right]^{\frac{1}{2}}.
\end{array}
\end{equation}
In particular, for $S_z=S=S_b-S_a$,
\begin{equation}
\begin{array}{l}
\displaystyle
g(S_{bz})=\langle S_a,S_a-S_b-S_{bz}; S_b,S_{bz}|S_a,S_b,S_a-S_b,S_a-S_b\rangle
= (-1)^{S_{bz}} \left[ \frac{(2S_b-2S_a+1)(2S_a)!(S_b+S_{bz})!}
{(2S_b+1)(2S_a-S_b+S_{bz})!} \right]^{\frac{1}{2}}.
\end{array}
\end{equation}

\end{widetext}

Furthermore, we can know the structure of  $|S_a,S_b,S,S\rangle$ in units of singlet pairs of atoms, using the discussions in Ref.~\cite{shi5}.

If $N_a+N_b-S$ is even,
\begin{widetext}
\begin{equation}
|S_a,S_b,S,S\rangle= \sum
A(Q_{1,1,0},Q_{2,0,0},Q_{0,2,0})
 {a_1^{\dagger}}^{Q_{1,0,1}}
{b_1^{\dagger}}^{Q_{0,1,1}}
{\Theta_{1,1,0}^{\dagger}}^{Q_{1,1,0}}
{\Theta_{2,0,0}^{\dagger}}^{Q_{2,0,0}} {\Theta_{0,2,0}^{\dagger}}^{Q_{0,2,0}}|0\rangle,
\label{g42}
\end{equation}
where $$\Theta_{1,1,0}^{\dagger}\equiv a_1^{\dagger}b_{-1}^{\dagger}
-a_0^{\dagger}b_{0}^{\dagger}+a_{-1}^{\dagger}b_{1}^{\dagger} $$ is the creation operator for a interspecies singlet pair~\cite{shi5}, $\Theta_{2,0,0}^{\dagger} \equiv \Theta_a^{\dagger}$ and $\Theta_{2,0,0}^{\dagger} \equiv \Theta_b^{\dagger}$ are creation operators for intraspecies  singlet pairs,
$Q_{1,0,1}$ and $Q_{0,1,1}$ are given by
\begin{equation}
\displaystyle
\left\{ \begin{array}{lcl}
Q_{1,0,1} & = & \frac{N_a-N_b+S}{2} + Q_{2,0,0} - Q_{0,2,0}, \\
Q_{0,1,1} & = & \frac{N_b-N_a+S}{2} + Q_{0,2,0} - Q_{2,0,0},
\end{array} \right. \label{q1}
\end{equation}

A special case is the singlet state under $N_a=N_b$, for which it is clear that $Q_{1,0,1}=Q_{0,1,1}=0$, that is in each term in the superposition, all atoms are paired up.

If $N_a+N_b-S$ is odd,
\begin{equation}
\begin{array}{cl}
\displaystyle
|S_a,S_b,S,S\rangle = &
\sum
A(Q_{1,1,0},Q_{2,0,0},Q_{0,2,0}) a_0^{\dagger}
 {a_1^{\dagger}}^{Q_{1,0,1}}
{b_1^{\dagger}}^{Q_{0,1,1}+1}
{\Theta_{1,1,0}^{\dagger}}^{Q_{1,1,0}}
{\Theta_{2,0,0}^{\dagger}}^{Q_{2,0,0}} {\Theta_{0,2,0}^{\dagger}}^{Q_{0,2,0}}|0\rangle \\
\displaystyle
&
+\sum A'(Q_{1,1,0},Q_{2,0,0},Q_{0,2,0}) {a_1^{\dagger}}^{Q_{1,0,1}+1}b_0^{\dagger}
{b_1^{\dagger}}^{Q_{0,1,1}}
{\Theta_{1,1,0}^{\dagger}}^{Q_{1,1,0}}
{\Theta_{2,0,0}^{\dagger}}^{Q_{2,0,0}} {\Theta_{0,2,0}^{\dagger}}^{Q_{0,2,0}}|0\rangle,
\end{array}
\label{g422}
\end{equation}
where $Q_{1,0,1}$ and $Q_{0,1,1}$ are given by
\begin{equation}
\displaystyle
\left\{ \begin{array}{lcl}
Q_{1,0,1} & = & \frac{N_a-N_b+S-1}{2} + Q_{2,0,0} - Q_{0,2,0}, \\
Q_{0,1,1} & = & \frac{N_b-N_a+S-1}{2} + Q_{0,2,0} - Q_{2,0,0}. \end{array} \right. \label{q2}
\end{equation}
In both (\ref{g42}) and (\ref{g422}),
the summations  are  over $Q_{1,1,0},$ $Q_{2,0,0}$ and $Q_{0,2,0}$.
The coefficients $A$ and $A'$ are determined by constraints
  \begin{equation}
\begin{array}{rcl}
S_a^2  |S_a,S_b, S,S\rangle & = & S_a(S_a+1)  |S_a,S_b, S,S\rangle, \\
S_b^2  |S_a,S_b, S,S\rangle & = & S_b(S_b+1)  |S_a,S_b, S,S\rangle.
\end{array} \label{sab}
\end{equation}
\end{widetext}
where $S_{\alpha}^2 = (\alpha_1^{\dagger}\alpha_1- \alpha_{-1}^{\dagger}\alpha_{-1})^2
+ 2(\alpha_0^{\dagger})^2\alpha_1\alpha_{-1} + 2 \alpha_1^{\dagger}\alpha_{-1}^{\dagger}\alpha_0^2 + 2\alpha_0^{\dagger} \alpha_0\alpha_1\alpha_1^{\dagger} + 2 \alpha_0^{\dagger} \alpha_0\alpha_{-1}\alpha_{-1}^{\dagger} $.

For a ground state which is a direct product of the states of individual species, one can resort to the composite structure of a single species of spin-1 atoms~\cite{hoyin}, and does not need to use the result here. For an entangled state of the form of $|S_a^m,S_b^m,|S_a^m-S_b^m|,|S_a^m-S_b^m|\rangle$, one can use the above result.

Now consider a particular  ground state  $|N_a,N_b,N_a-N_b,N_a-N_b\rangle$. As now $N_a+N_b-S$ is even, we know that  $|N_a,N_b,N_a-N_b,N_a-N_b\rangle$ is given by Eq.~(\ref{g42}), with \begin{equation}
\displaystyle
\left\{ \begin{array}{lcl}
Q_{1,0,1} & = & N_a-N_b + Q_{2,0,0} - Q_{0,2,0}, \\
Q_{0,1,1} & = & Q_{0,2,0} - Q_{2,0,0}.
\end{array} \right.
\end{equation}

\section{summary \label{summary}}

Under the single spatial mode approximation for each atom,
the  many-body Hamiltonian of a mixture of two species, labeled as $a$ and $b$, of spin-1 atoms with both intraspecies  and interspecies spin exchanges can be reduced to that of two coupled giant spins, with the intraspecies  and interspecies spin coupling strengths $c^a$, $c^b$ and $c^{ab}$, as given in (\ref{spinham}). As listed in Table~\ref{table1}, we have determined the ground states in nearly all parameter regimes, except a small patch where there are too many possibilities, by minimizing the energy under various conditions. These states can be rewritten in terms of bosonic degree of freedom, as discussed in Sec.~\ref{comp}. The states can always be written as superpositions of various configurations of intraspecies  and interspecies spin singlet pairs, together with some individual atoms with z-component spin $1$, the total number of which is $S_z$. For a many-body singlet ground state, all the atoms are paired up in each configuration.

The many-body singlet ground states in two regimes deserve particular attention. One is the non-degenerate singlet ground state $|N_b,N_b,0,0\rangle$ in the regime $c^{ab} >0$, $c^{b} \leq c^{ab}$ and  $\frac{2N_bc^{ab}}{2N_b+1} \leq c^a \leq c^{ab}$ (excluding $c^a=c^b=c^{ab}$)  plus  $c^b \leq  0$ and $c^{ab} < c^{a} \leq \frac{2(N_b+1)c^{ab}}{2N_b+1}$. This singlet state exists even in the generic case of $N_a \geq N_b$.   In the special case of $N_a=N_b$, its regime only expands to $c^{b} \leq c^{ab}$ and  $ c^a \leq c^{ab}$ plus  $c^b \leq  0$, and $c^{ab} < c^{a}\leq \frac{2(N_b+1)c^{ab}}{2N_b+1}$ and plus $c^{b} \leq \frac{2(N_a+1)c^{ab}}{2N_a+1}$ while $c^a\leq 0$.

The other regime of many-body singlet states is $c^a=c^b=c^{ab}>0$, where the ground state can be any $|S_b^m,S_b^m,0,0\rangle$ with a legitimate value of $S_b^m$. Therefore in the generic case of $N_a\geq N_b$,  in the regime $c^{ab} >0$, $c^{b} \leq c^{ab}$ and  $\frac{2N_bc^{ab}}{2N_b+1} \leq c^a \leq c^{ab}$ (including $c^a=c^b=c^{ab}$)  plus  $c^b \leq  0$ and $c^{ab} < c^{a} \leq \frac{2(N_b+1)c^{ab}}{2N_b+1}$, we always have $S=S_z=0$, even though the magnetic field is absolutely zero. This robustness is useful for experiments.

At $c^a=c^b=c^{ab}$, there is a continuous quantum phase transition from the global singlet state $|N_b,N_b,0,0\rangle$ to the product of two single-species singlet states  $|0,0\rangle_a|0,0\rangle_b$, which is the ground state in the regime $c^a >0$, $c^b>0$ and $c^ac^b >(c^{ab})^2$.

Focusing on the maximally polarized states with $S_z=S^m$, as can be picked out by an infinitesimal magnetic field, we have described the phase-diagrams on $c^a-c^b$ planes for zero, negative and positive $c^{ab}$, respectively, thus obtaining the three-dimensional phase diagram in  $c^a-c^b-c^{ab}$ parameter space.

For $c^{ab}=0$ (FIG.~\ref{pd3}), the four quadrants correspond to the products of two antiferromagnetic states $|0,0\rangle_{a}|0,0\rangle_{b}$, of a ferromagnetic state and an antiferromagnetic state $|N_a,N_a\rangle_{a}|0,0\rangle_{b}$ and $|0,0\rangle_a|N_b,N_b\rangle_{b}$, and  of  two  ferromagnetic states $|N_a,N_a\rangle_{a}|N_b,N_b\rangle_{b}$, respectively. Each ground state is a distinct quantum phase, with the $c^a$ and $c^b$ axes as the phase boundaries, on which there are degenerate ground states, two of which belong to the two bordering phases, respectively.

Another kind of phase boundaries appear in the case of  $c^{ab}<0$ (FIG.~\ref{pd1}). With the increase of $|c^{ab}|$, there grows a surface $c^b = \frac{2N_a c^{ab}}{2N_b+1}$   in  the part  $ c^a \geq 0 $  and
a surface $c^a = \frac{2N_b c^{ab}}{2N_a+1}$ in the part $c^b \geq 0 $. These two surfaces,  together with $c^b-c^{ab}$ plane, ($c^a=0$)  in the part
$c^b \geq  \frac{2N_a c^{ab}}{2N_b+1}$,  and $c^a-c^{ab}$ plane ($c^b=0$)  in the part $c^a\geq  \frac{2N_b c^{ab}}{2N_a+1}$, act as the phase boundary separating  $|0,0\rangle_{a}|0,0\rangle_{b}$ from the other phase, to which the phase boundary itself also belongs.

The latter  phase for  $c^{ab}<0$  consists of five ground states, which are continuously connected on four boundaries,  $c^a = \frac{-2N_bc^{ab}}{2N_a+1}$ and $c^a= -2N_bc^{ab}$ for $c_b < 0$, and  $c^b=\frac{-2N_ac^{ab}}{2N_b+1}$ and $c^b=-2N_ac^{ab}$ for $c^a <0$. These four boundaries of continuous connection  converge respectively to negative halves of $c^a$ and $c^b$ axes as $c^{ab} \rightarrow 0$. Especially, for a given $c^{ab}<0$, the crossover regime of  $|N_a,N_a\rangle_a|n_1,n_1\rangle_b$  disappears as $c^{ab} \rightarrow 0$. Consequently,  $|N_a,N_a\rangle_a|N_b,N_b\rangle_b$ and $|N_a,N_a\rangle_a|0,0\rangle_b$, which can be continuously related through $|N_a,N_a\rangle_a|n_1,n_1\rangle_b$, become discontinuous neighbors as $c^{ab} \rightarrow 0$.
Similarly, the crossover regime of $|n_2,n_2\rangle_a|N_b,N_b\rangle_b$  also disappears as $c^{ab} \rightarrow 0$. Consequently,  $|N_a,N_a\rangle_a|N_b,N_b\rangle_b$ and $|0,0\rangle_a|N_b,N_b\rangle_b$, which can be continuously related through $|n_2,n_2\rangle_a|N_b,N_b\rangle_b$, become discontinuous neighbors  as $c^{ab} \rightarrow 0$.  This conforms with the $c^a-c^b$ phase diagram for $c^{ab}=0$, in which the negative $c^a$ axis and the negative $c^b$ axis are phase boundaries.

With the increase of $c^{ab}$ on the positive direction, there also grow a number of boundaries where the ground states are continuously connected, consequently the ground states in the second, third and fourth octants, plus the part of first octant with $c^a \leq c^{ab}$ and $c^b \leq c^{ab}$ (excluding $c^a=c^b=c^{ab}$) are all continuously connected to be a single quantum phase  (FIG~\ref{pd2}).  For $c^a \leq 0$, there are two boundaries of continuous connection  $c^b= \frac{2(N_a+1)c^{ab}}{2N_b+1}$ and $c^b= 2(N_a+1) c^{ab}$, between which is the  crossover regime of  ground states $|N_a,n_4,N_a-n_4,N_a-n_4\rangle$, which is continuously connected downwards with
$|N_a,N_b,N_a-N_b,N_a-N_b\rangle$ and  upwards to $|N_a,N_a\rangle_a|0,0\rangle_b$.    In case $N_a >N_b$, there are two  boundaries of continuous connection $c^a= \frac{2N_bc^{ab}}{2N_a+1}$ and $c^a= \frac{2N_bc^{ab}}{2N_b+1}$ for $c^b\leq c^{ab}$, which disappear in case $N_a=N_b$.  These two boundaries define the crossover regime of
$|n_2,N_b,n_2-N_b,n_2-N_b\rangle$,  which is continuously connected leftwards with $|N_a,N_b,N_a-N_b,N_a-N_b\rangle$, and rightwards with
$|N_b,N_b,0,0\rangle$.  For $c^b \leq 0$, there are two  boundaries of continuous connection  $c^a=\frac{2(N_b+1)c^{ab}}{2N_b+1}$ and $2(N_b+1) c^{ab}$, between which is the crossover regime of
$|n_3,N_b,N_b-n_3,N_b-n_3\rangle$, which  is continuously connected leftwards with $|N_b,N_b,0,0\rangle$, and rightwards with   $|0,0\rangle_a|N_b,N_b\rangle_b$. When $c^{ab}\rightarrow 0$, the crossover regimes all tend to disappear, therefore $|N_a,N_a\rangle_a|0,0\rangle_b$ and $|0,0\rangle_a|N_b,N_b\rangle_b$ tend to occupy the second and fourth quadrant of the $c^a-c^b$ cross-section, as for $c^{ab}=0$. But unlike the case of $c^{ab} <0$, there is a discontinuity in the fourth quadrant between $ |N_a,N_b,N_a-N_b,N_a-N_b\rangle$
for $c^{ab} >0$ and $|N_a,N_a\rangle_a|N_b,N_b\rangle_b$ for $c^{ab} =0$.
In the first octant, with the increase of $c^{ab}$, the boundary of $|0,0\rangle_a|0,0\rangle_b$ grows from the positive $c^a$ and $c^b$ axes to $c^a c^b = (c^{ab})^2$, $c^a >0$ and $c^b >0$.

Interspecies spin coupling $c^{ab}$ provides an extra parameter controlling quantum phase transitions. We have given two examples. First, the  discontinuity between ground states for $c^{ab} > 0$ and $c^{ab} \leq  0$ when $c^a \leq 0$ and $c^b \leq 0$ implies a first order quantum phase transition in varying the sign of $c^{ab}$. Second, by tuning $c^{ab}$ from zero to negative and then back to zero, the first order quantum phase transition between negative and positive values of an intraspecies  spin coupling can be circumvented.

With the experimental experiences on single-species spinor gases~\cite{spinor},  and on scalar mixtures~\cite{myatt}, where interspecies spin exchange has been observed as a disadvantage, and with the development of heteronuclear Feshbach resonance~\cite{chin}, future experimental realization of a spinor mixture with interspecies spin exchange, with the ground states described here, is expected.

\acknowledgments

This work was supported by the National Science Foundation of China (Grant No. 11074048), the Shuguang Project (Grant No. 07S402) and the Ministry of Science and Technology of China (Grant No. 2009CB929204).

\appendix

\section{$S_a^m,S_b^m$ and $S^m$ in the case $c^{ab} < {0}$}

In this appendix, we find  $S_a^m,S_b^m$ and $S^m$, in which $E$ is minimal, in  parameter regimes with $c^{ab} < 0$ and in the absence of a magnetic field.
In the discussion, $E$ always represents the energy as low as can be determined in the regime under discussion, i.e. the meaning of $E$ keeps updating.

With $c^{ab} < 0$, $E$ is minimal when $S=S_a+S_b$. Hence
\begin{equation}
E=\frac{c^{a}}{2}S_a(S_a+1)
+\frac{c^{b}}{2}S_b(S_b+1)+c^{ab}S_aS_b. \label{afirst}
\end{equation}
Thus
\begin{equation}
\frac{\partial{E}}{\partial{S_a}}=
c^{a}S_a+c^{ab}S_b+\frac{c^{a}}{2},
\end{equation}
\begin{equation}
\frac{\partial{E}}{\partial{S_b}}=
c^{b}S_b+c^{ab}S_a+\frac{c^{b}}{2}.
\end{equation}
Several different cases are considered in the following.

\subsection{$c^{a}\leq 0$, $c^{b} \leq 0$ \label{casea1}}

In this regime, of which regime I in   Ref.~\cite{shi5} is a subset,  $\frac{\partial{E}}{\partial{S_a}}<0$,
$\frac{\partial{E}}{\partial{S_b}}<0$, hence $S^m_a=N_a$,
$S^m_b=N_b, S^m=N_a+N_b$.

\subsection{$c^{a}\leq 0$, $c^{b}>0$ \label{casea2}}

In this regime,
$\frac{\partial{E}}{\partial{S_a}}<0$, hence $S_a^m=N_a$,  \begin{equation}
E(N_a,S_b) =\frac{c^{b}}{2}S_b(S_b+1)+c^{ab}N_aS_b+const.
\end{equation}

\begin{figure}
\begin{center}
\scalebox{0.6}{\includegraphics[126,538][424,750]{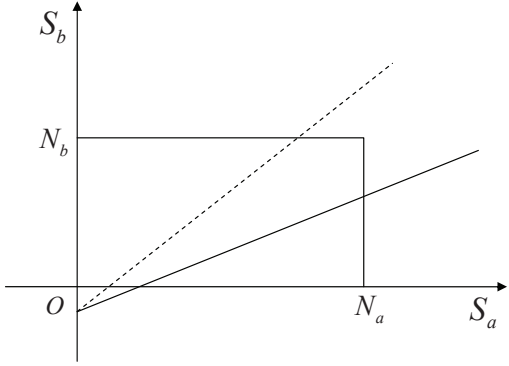}}
\end{center}
\caption{\label{figa}  Possible values of $S_a$ and $S_b$ are within the rectangular defined by $0 \leq S_a \leq N_a$ and  $0 \leq S_b \leq N_b$ on $S_a$-$S_b$ plane. The dashed line is the stationary line $\frac{\partial{E}}{\partial{S_b}}=0$ in the case  $c^{ab} < 0$ and $0 <c^{b} \leq -\frac{2N_ac^{ab}}{2N_b+1}$. The solid line is the  stationary line in the case $c^{ab} < 0$ and  $ -\frac{N_a c^{ab}}{N_b+1/2} \leq c^{b} \leq  -2 N_a c^{ab}$. }
\end{figure}

We represent all the values of $S_a$ and $S_b$ as points $(S_a, S_b)$ within the rectangular defined by $0 \leq S_a \leq N_a$ and  $0 \leq S_b \leq N_b$ on
$S_a$-$S_b$ plane (FIG.~\ref{figa}).
$\frac{\partial{E}}{\partial{S_b}}=0$  defines a stationary line.
The points above this line satisfy $\frac{\partial{E}}{\partial{S_b}}>0$, while the points below it satisfy  $\frac{\partial{E}}{\partial{S_b}}<0$.  One can see three possibilities.

\subsubsection{ $0 <c^{b} \leq -\frac{2N_ac^{ab}}{2N_b+1}$}

The stationary
line, depicted as the dashed line, crosses with the line $S_b=N_b$. Hence all points with $S_a=N_a$
satisfy $\frac{\partial{E}}{\partial{S_b}} \leq 0$. Consequently
$S_a^m=N_a$, $S_b^m=N_b$, $S^m=N_a+N_b$. Note that this regime can
be combined with case \ref{casea1}, where the solution is the  same.

\subsubsection{  $ -\frac{N_a c^{ab}}{N_b+1/2} \leq c^{b} \leq  -2 N_a c^{ab}$}

The stationary line, depicted as the solid line in FIG.~\ref{figa}, crosses with the line $S_a=N_a$. The crossing point gives the minimal energy.
Hence $S_a^m=N_a$, $S_b^m=n_1$, with
\begin{equation}
n_1 \equiv {\rm  Int}[\frac{|c^{ab}|}{c^{b}}N_a-\frac{1}{2}],
\end{equation}
where ${\rm  Int}(x)$ represents the integer closest to $x$ and in the legitimate range  of $S_b$; here $0\leq n_1 \leq N_b$.  $S^m=N_a + n_1$.

\subsubsection{ $c^{b} \geq -2 N_a c^{ab}$}

If $c^{b} > -2 N_a c^{ab}$, all points $(S_a, S_b)$ in the rectangular satisfy $\frac{\partial{E}}{\partial{S_b}} > 0$. Therefore $S_b^m =0$, $S_a^m=S^m =N_a$. The same solution is obtained if $c^{b} = -2 N_a c^{ab}$.

\subsection{$c^{a}>0$, $c^{b}\leq 0$ \label{casea3} }

One simply exchanges the subscripts or superscripts  $a$ and $b$ in the preceding subcase. Thus there are also three possibilities.

\subsubsection{ $0<c^{a} \leq -\frac{N_b c^{ab}}{N_a+1/2}$}

$S_a^m=N_a$, $S_b^m=N_b$, $S^m=N_a+N_b$. This regime can also be combined with \ref{casea1}, with the same solution.

\subsubsection{ $ -\frac{N_b c^{ab}}{N_a+1/2} \leq c^{a} \leq  -2 N_b c^{ab}$.}

$S_a^m=n_2$, with
\begin{equation}
n_2 \equiv {\rm  Int}[\frac{|c^{ab}|}{c^{a}}N_b-\frac{1}{2}],
\end{equation}
with $0\leq n_2 \leq N_a$, $S_b^m=N_b$,  $S^m= n_2 + N_b$.

\subsubsection{$c^{a} \geq -2 N_b c^{ab}$}

 $S_a^m = 0$, $S_b^m =S^m = N_b$

\subsection{ $c^{a}>0$, $0<c^{b}<-2 N_a c^{ab} $,
$c^{a}c^{b} \geq (c^{ab})^2$ \label{sectiona} }

First we consider $0<c^{b}\leq -\frac{2N_ac^{ab}}{2N_b+1}$.
Then the  stationary
line $\frac{\partial{E}}{\partial{S_b}}=0$ crosses with the line
$S_b=N_b$, as shown as the dashed line in FIG~\ref{figa}. The crossing points with $S_b=0$ and
$S_b=N_b$ divide the rectangular to three regions, for which the
minima of $E$ are
\begin{widetext}
\begin{equation}
E=\left\{ \begin{array}{ll}
 \frac{c^{a}}{2}S_a(S_a+1),  &  \text{if} \quad 0\leq{S_a} \leq -\frac{c^{b}}{2c^{ab}}, \\
 \frac{c^{a}c^{b}-(c^{ab})^2}{2c^{b}}S_a^2+\frac{c^{a}-c^{ab}}{2}S_a-
 \frac{c^{b}}{8}, & \text{if} \quad -\frac{c^{b}}{2c^{ab}}\leq{S_a} \leq -\frac{(N_b+1/2)c^{b}}{c^{ab}}, \\
 \frac{c^{a}}{2}S_a(S_a+1)+c^{ab}N_bS_a+\frac{c^{b}}{2}N_b(N_b+1),
 & \text{if} \quad -\frac{(N_b+1/2)c^{b}}{c^{ab}}\leq{S_a}\leq{N_a}.
 \end{array} \right. \label{e31}
 \end{equation}
 \end{widetext}
$S_b=0$ in the first interval of $S_a$. In the second interval, $S_a$ and $S_b$ satisfy $\partial E/\partial S_b =0$. In the third interval, $S_b=N_b$.
In each of these three intervals, $E$ increases as $S_a$ increases. Therefore, $S_a^m=S_b^m=S^m=0$.

Now we consider $-\frac{2N_ac^{ab}}{2N_b+1}<c^{b}<-2 N_a c^{ab}$.
Then the  stationary
line $\frac{\partial{E}}{\partial{S_b}}=0$ crosses with the line
$S_a=N_a$, as shown as the solid line in FIG.~\ref{figa}. The crossing points with $S_b=0$ and
$S_a=N_a$ divide the rectangular to two regions.
The minima of $E$ are given by
\begin{widetext}
\begin{equation}
E=\left\{ \begin{array}{ll}
 \frac{c^{a}}{2}S_a(S_a+1),  &  \text{if} \quad 0\leq{S_a}\leq -\frac{c^{b}}{2c^{ab}}, \\
 \frac{c^{a}c^{b}-(c^{ab})^2}{2c^{b}}S_a^2+\frac{c^{a}-c^{ab}}{2}S_a-
 \frac{c^{b}}{8}, & \text{if} \quad -\frac{c^{b}}{2c^{ab}}\leq{S_a}\leq{N_a}.
 \end{array} \right. \label{e32}
 \end{equation}
 \end{widetext}
$S_b=0$ in the first interval of $S_a$. In the second region, $S_a$ and $S_b$ satisfy $\partial E/\partial S_b =0$.
Again it is found in each region, $E$ increases as $S_a$ increases, therefore  $S_a^m=S_b^m=S^m=0$.

Therefore, throughout  this parameter regime, we have $S_a^m=S_b^m=S^m=0$.

\subsection{ $c^{b}>0$, $0<c^{a}<-2 N_b c^{ab} $,
$c^{a}c^{b} \geq (c^{ab})^2$  }

Exchanging labels $a$ and $b$ in the preceding subcase, we obtain $S_a^m = S_b^m =0$.

\subsection{ $c^{a} > 0$, $ c^b > -2 N_a c^{ab}$ \label{casea6} }

All points satisfy $\partial E/\partial S_b > 0$. Hence $S_b^m=0$. Consequently, one also obtains, from Eq.~(\ref{afirst}), $S_a^m =0$.

\subsection{ $c^{b} > 0$, $ c^a > -2 N_b c^{ab}$ \label{casea7} }

Exchanging labels $a$ and $b$ in the preceding subcase, we have $S_a^m = S_b^m =0$

\subsection{ $c^{a}>0$, $-\frac{2N_ac^{ab}}{2N_b+1} \leq c^b \leq -2 N_a c^{ab}$, $c^{a}c^{b} < (c^{ab})^2$ }

Now $E$ is as given in (\ref{e32}). With $c^{a}c^{b} < (c^{ab})^2$, $\partial E/\partial S_a > 0$ in the the second region. Consequently $S_a^m=S^m_b=S^m=0$.

\subsection{ $c^{b}>0$, $-\frac{2N_bc^{ab}}{2N_a+1} \leq c^a \leq -2 N_b c^{ab}$, $c^{a}c^{b} < (c^{ab})^2$ }

Exchanging labels $a$ and $b$ in the preceding subcase,  we also obtain $S_a^m=S^m_b=S^m=0$.

\subsection{ $0< c^{a} \leq -\frac{2N_bc^{ab}}{2N_a+1}$, $0<c^{b}\leq -\frac{2N_ac^{ab}}{2N_b+1}$,
$c^{a}c^{b} <  (c^{ab})^2$  \label{casea10}  }

$E$ is as given in (\ref{e31}). In the first interval of $S_a$, the minimum of $E$ is at $S_a=0$. In the second interval, the minimum is either at  $S_a=-\frac{(N_b+1/2)c^{b}}{c^{ab}}$, which also belongs to the first interval,  or at $S_a = -\frac{c^{b}}{2c^{ab}}$, which also belongs to the third region.   From
$c^{a}\leq -\frac{N_bc^{ab}}{N_b+1/2}$, we have $n_2 \geq N_a$. Consequently, the minimum in the third interval is at $S_a=N_a$, moreover, it is straightforward to see $E(S_a=N_a) < E(0)$. Therefore, $S_a^m=N_a$, $S_b^m = N_b$, $S^m=N_a+N_b$.

\section{ $c^{ab}>{0}$}

In this appendix, we find out $S_a^m,S_b^m$ and $S^m$, in which $E$ is minimal, in  parameter regimes with $c^{ab} > 0$ and in absence of the magnetic field.  Without loss of generality,  assume  $N_a \geq N_b$.

\subsection{$c^a=c^b=c^{ab}$}

In this case, $S_a$ and $S_b$ both disappear in $E$. Hence $E$ is minimal when $S=0$. Thus $S_a^m=S_b^m$ for any legitimate value of $S_b^m$, i.e. $S_a^m=S_b^m=0,\cdots,N_b$.

In any other cases, as considered in the following,
$E$ is minimal when $S=|S_a-S_b|$.

\subsection{$c^{a} \leq c^{ab}$, $c^{b} \leq c^{ab}$, except $c^a=c^b=c^{ab}$ }

\begin{figure}
\begin{center}
\scalebox{0.6}{\includegraphics[120,520][404,760]{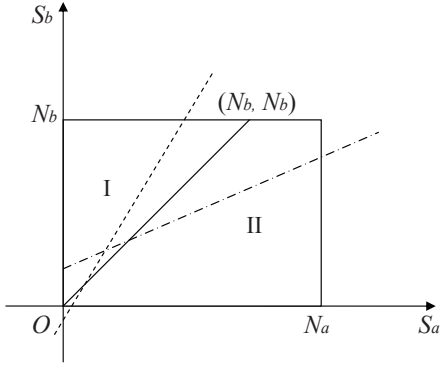}}
\end{center}
\caption{\label{figb} Possible values of $S_a$ and $S_b$ are within the rectangular defined by $0 \leq S_a \leq N_a$ and  $0 \leq S_b \leq N_b$ on $S_a$-$S_b$ plane. The solid line represents $S_a=S_b$, above which is region I while below which is region II.  The dashed line is the stationary line $\frac{\partial{E}}{\partial{S_b}}=0$ in region I in the case $c^{a}>c^{ab}>c^{b}>0$. The broken line is the  stationary line in region II in the case  $c^{b}>c^{ab}>c^{a}>0$.   }
\end{figure}

As shown in
FIG.~{\ref{figb}, we divide the whole region of $S_a$ and $S_b$ to two regions
by the line $S_b =S_a$. In region I, $E$ is minimal when
$S=S_b-S_a$, i.e.
\begin{equation}
E=\frac{c^{a}}{2}S_a(S_a+1)+
\frac{c^{b}}{2}S_b(S_b+1)-c^{ab}S_aS_b-c^{ab}S_a. \label{b1}
\end{equation}
Hence
\begin{equation}
\frac{\partial{E}}{\partial{S_a}}=c^{a}S_a-c^{ab}S_b+\frac{c^{a}}{2}-c^{ab}<0.
\end{equation}
Therefore, the minimum of $E$ in region I lies on the border line
$S_b=S_a$, on which
\begin{equation}
E=\frac{c^{a}+c^{b}-2c^{ab}}{2}S_a(S_a+1).
\end{equation}
which reaches its minimum at
$(N_b,N_b)$, as $c^{a}+c^{b}-2c^{ab}<0$.

In region II, $S=S_a-S_b$,
\begin{equation}
E=\frac{c^{a}}{2}S_a(S_a+1)+\frac{c^{b}}{2}S_b(S_b+1)-
c^{ab}S_aS_b-c^{ab}S_b. \label{a9}
\end{equation}
Hence
\begin{equation}
\frac{\partial{E}}{\partial{S_b}}=c^{b}S_b-c^{ab}S_a
+\frac{c^{b}}{2}-c^{ab}<0. \label{partialb}
\end{equation}
Therefore, the minimum of $E$ in region II lies on the border line
$S_a=S_b$ for $0\leq S_a \leq N_b$, and on the line $S_b=N_b$ for
$N_b \leq S_a\leq N_a$.

Therefore, the minimum of $E$ in the whole rectangular must lie on
the line interval $S_b=N_b$, $N_b \leq S_a\leq N_a$, on which
\begin{equation}
E=\frac{c^{a}}{2}S_a(S_a+1)-c^{ab}N_bS_a+const. \label{esb}
\end{equation}
There are three possibilities.

\subsubsection{ $c^{a} \leq  \frac{2N_bc^{ab}}{2N_a+1}$}

This includes the range that $c^a < 0 $.  $S_a^m=N_a$,
$S_b^m=N_b$,  $S^m = N_a - N_b$.

\subsubsection{  $\frac{2N_bc^{ab}}{2N_a+1} \leq c^a \leq \frac{2N_bc^{ab}}{2N_b+1} $}

We have
$S_a^m=n_2$,
$S_b^m=N_b$, $S^m = n_2-N_b$.

\subsubsection{  $\frac{2N_bc^{ab}}{2N_b+1} \leq c^a \leq c^{ab} $}

We have
$S_a^m=S_b^m=N_b$, $S^m =0$.

\subsection{$c^{a}>c^{ab}>c^{b}>0$, $c^{a}c^{b}>(c^{ab})^2$ \label{case21}}

In region II,  $\frac{\partial{E}}{\partial{S_b}}<0$, as shown in
Eq.~(\ref{partialb}). Hence  the minimum of $E$  lies on the border
line $S_a=S_b$ for $0\leq S_a \leq N_b$, and on the line $S_b=N_b$
for $N_b \leq S_a\leq N_a$. When $S_b=N_b$, $E$ is given by
(\ref{esb}), for which we now have
\begin{equation}
\frac{\partial{E}}{\partial{S_a}}=c^{a}S_a-c^{ab}N_b+\frac{c^{a}}{2}\geq
 (c^{a}-c^{ab})N_b+\frac{c^{a}}{2}>0.
 \end{equation}
Therefore in the line interval  $S_b=N_b$
for $N_b \leq S_a\leq N_a$, the minimum of $E$ locates at $(N_b, N_b)$.
Consequently the minimum of  $E$ in the whole rectangular lies in region I, in which $E$ is given in (\ref{b1}), thus
\begin{equation}
\frac{\partial{E}}{\partial{S_b}}
=c^{b}S_b-c^{ab}S_a+\frac{c^{b}}{2}. \label{b8}
\end{equation}
$\frac{\partial{E}}{\partial{S_b}}=0$ defines a stationary line
which crosses with the lines
 $S_a=S_b$ and $S_b=N_b$ at $(\frac{c^{b}}{2(c^{ab}-c^{b})}, \frac{c^{b}}{2(c^{ab}-c^{b})})$
 and $(\frac{c^{b}(N_b+1/2)}{c^{ab}}, N_b)$ respectively, depicted as the dashed line in
FIG.~{\ref{figb}.

If $\frac{c^{b}}{c^{ab}} \leq \frac{N_b}{N_b+1/2}$, the crossing point with $S_b=N_b$ lies in region I, then the minima  of $E$ can be found to be
 \begin{widetext}
 \begin{equation}
 E= \left\{ \begin{array}{ll}
 \frac{c^{a}+c^{b}-2c^{ab}}{2}S_a(S_a+1), & \text{if} \quad 0\leq{S_a}<\frac{c^{b}}{2(c^{ab}-c^{b})} \\
 \frac{c^{a}c^{b}-(c^{ab})^2}{2c^{b}}S_a^2+
 \frac{c^{a}-c^{ab}}{2}S_a-\frac{c^b}{8}, & \text{if} \quad \frac{c^{b}}{2(c^{ab}-c^{b})}\leq{S_a} \leq \frac{c^{b}(N_b+1/2)}{c^{ab}} \\
 \frac{c^{a}}{2}S_a(S_a+1)-c^{ab}N_bS_a-c^{ab}S_a
 +\frac{c^b}{2}N_b(N_b+1),
 & \text{if} \quad \frac{c^{b}(N_b+1/2)}{c^{ab}}<{S_a}\leq{N_b}
 \end{array} \right.  \label{a15}
 \end{equation}
Thus
 \begin{equation}
 \frac{\partial{E}}{\partial{S_a}}=\left\{ \begin{array}{ll}
 (c^{a}+c^{b}-2c^{ab})(S_a+\frac{1}{2}), & \text{if} \quad 0\leq{S_a}<\frac{c^{b}}{2(c^{ab}-c^{b})} \\
 \frac{c^{a}c^{b}-(c^{ab})^2}{c^{b}}S_a+\frac{c^{a}-c^{ab}}{2}, & \text{if} \quad \frac{c^{b}}{2(c^{ab}-c^{b})}\leq{S_a} \leq \frac{c^{b}(N_b+1/2)}{c^{ab}} \\
 c^{a}(S_a+\frac{1}{2})-c^{ab}N_b-c^{ab}, & \text{if} \quad \frac{c^{b}(N_b+1/2)}{c^{ab}} < {S_a}\leq{N_b}
 \end{array} \right.
 \end{equation}
 \end{widetext}
As $c^{a}c^{b}>(c^{ab})^2$,
$c^{a}+c^{b}>2\sqrt{c^{a}c^{b}}> 2c^{ab}$, we
find that $\frac{\partial{E}}{\partial{S_a}}>0$ in all the three
intervals. Consequently  $S_a^m=S_b^m=S^m=0$.

If $\frac{c^{b}}{c^{ab}} > \frac{N_b}{N_b+1/2}$,  then the
$\frac{\partial{E}}{\partial{S_b}}>0$ in the whole region II. Then the minima in the whole region I are given by the first expression in (\ref{a15}). One again obtains  $S_a^m=S_b^m=S^m=0 $.

\subsection{$c^{b}>c^{ab}>c^{a}>0$, $c^{a}c^{b}>(c^{ab})^2$ \label{case22}}

One still uses FIG.~\ref{figb}. In region I,  $\frac{\partial{E}}{\partial{S_b}} >0$, hence  the minimum of $E$  lies on the border
line $S_a=S_b$.
Thus the minimum of  $E$ in the whole rectangular lies in region II, in which $E$ is given in (\ref{a9}), and thus
\begin{equation}
\frac{\partial{E}}{\partial{S_a}}
=c^{a}S_a-c^{ab}S_b+\frac{c^{a}}{2}.
\end{equation}
$\frac{\partial{E}}{\partial{S_a}}=0$ defines a stationary line, shown as the broken line in  FIG.~\ref{figb},
which crosses with the line
 $S_a=S_b$ at $(\frac{c^{a}}{2(c^{ab}-c^{a})}, \frac{c^{a}}{2(c^{ab}-c^{a})})$.

If $\frac{c^{a}}{c^{ab}} \geq \frac{ N_b}{N_a+1/2}$, the line
$\frac{\partial{E}}{\partial{S_a}}=0$  also crosses with the boundary $S_b=N_b$ . Then the minima of E can be found to be
 \begin{widetext}
 \begin{equation}
 E= \left\{ \begin{array}{ll}
 \frac{c^{a}+c^{b}-2c^{ab}}{2}S_a(S_a+1), & \text{if} \quad 0\leq{S_a}<\frac{c^{a}}{2(c^{ab}-c^{a})} \\
 \frac{ c^{a}}{2}(\frac{c^{a}c^{b}}{(c^{ab})^2}-1)S_a^2+
 c^a(-1+\frac{c^ac^{b}}{2(c^{ab})^2}+
 \frac{c^{b}}{2c^{ab}})S_a+ \frac{c^a}{2}(\frac{c^ac^b}{4(c^{ab})^2}+\frac{c^b}{2c^{ab}}-1), & \text{if} \quad \frac{c^{a}}{2(c^{ab}-c^{a})}\leq{S_a}
 \leq \frac{c^{ab}}{c^{a}}N_b-\frac{1}{2}\\
 \frac{c^{a}}{2}S_a(S_a+1)-c^{ab}N_bS_a+\frac{c^b}{2}N_b(N_b+1)-c^{ab}N_b,
 & \text{if} \quad \frac{c^{ab}}{c^{a}}N_b-\frac{1}{2} < {S_a}\leq{N_a}
 \end{array} \right.  \label{a17}
 \end{equation}
 \end{widetext}
With $c^{a}c^{b}>(c^{ab})^2$, it is found that $\frac{\partial{E}}{\partial{S_a}}>0$ in all the three
intervals. Consequently  $S_a^m=S_b^m=S^m=0$.

If $\frac{c^{a}}{c^{ab}} <\frac{ N_b}{N_a+1/2}$, the stationary line $\frac{\partial{E}}{\partial{S_a}}=0$  crosses with the boundary $S_a=N_a$. Then the minima of $E$ in region II is given by the first two expressions in (\ref{a17}), with the second interval of $S_a$ extended to $N_a$. One obtains the same result $S_a^m=S_b^m=S^m=0$.

\subsection{$c^{a} > c^{ab}$, $c^{b} > c^{ab}$}

This regime has been discussed in
\cite{shi5}, with result  $S_a^m=S_b^m=S^m=0$.

The cases \ref{case21}, \ref{case22} and the present one can be written altogether as
$c^{a}c^{b}>(c^{ab})^2$, with the same result
$S_a^m=S_b^m=S^m=0$.

\subsection{$c^{a}> c^{ab}>0$, $c^b \leq  0$}

The reasoning given in case \ref{case21} that the minimum of $E$ must lie in region I is still valid. Using (\ref{b8}), now $\partial E/\partial S_b < 0$, as $c^b \leq 0$. Therefore $S_b^m=N_b$, for which
\begin{equation}
E=\frac{c^{a}}{2}S_a(S_a+1)-
c^{ab}N_b S_a-c^{ab}S_a+const, \label{a6}
\end{equation}
with $0\leq S_a \leq N_b$.  There are three possibilities.

\subsubsection{$c^{ab} < c^{a} \leq \frac{2(N_b+1)c^{ab}}{2N_b+1}$, $c^b \leq 0$}

$S_a^m = N^b$, $S_b^m=N_b$, $S^m = 0$.

\subsubsection{$  \frac{2(N_b+1)c^{ab}}{2N_b+1} \leq c^{a} \leq 2(N_b+1) c^{ab}$, $c^b \leq 0$}

$S_a^m = n_3$, where
\begin{equation}
n_3 \equiv {\rm  Int}[\frac{c^{ab}(N^b+1)}{c^a} -\frac{1}{2}],
\end{equation}
with $0 \leq n_3 \leq N_b$.  $S_b^m=N_b$, $S^m = N_b -n_3$.

\subsubsection{$  c^{a} \geq 2(N_b+1) c^{ab}$, $c^b \leq 0$}

$S_a^m=0$, $S_b^m=S^m=N_b$.

\subsection{$c^{b}> c^{ab}>0$, $c^a \leq  0$}

As $c^{a}< c^{ab}$, the reasoning in  case \ref{case22} that
the minimum of $E$ must lie in region II is still valid.
Then $\partial E/\partial S_a < 0$. Therefore $S_a^m=N_a$, for which
\begin{equation}
E=\frac{c^{b}}{2}S_b(S_b+1)-
c^{ab}N_a S_b-c^{ab}S_b+const. \label{a62}
\end{equation}
 There are three possibilities.

\subsubsection{$c^{ab} < c^{b} \leq \frac{2(N_a+1)c^{ab}}{2N_b+1}$, $c^a \leq 0$}

$S_a^m = N^a$, $S_b^m=N_b$, $S^m = N_a-N_b$.

\subsubsection{$  \frac{2(N_a+1)c^{ab}}{2N_b+1} \leq c^{b} \leq 2(N_a+1) c^{ab}$, $c^a \leq 0$}

$S_a^m=N_a$,
$S_b^m = n_4$, where
\begin{equation}
n_4 \equiv {\rm  Int}[\frac{(N^a+1)c^{ab}}{c^b} -\frac{1}{2}],
\end{equation}
with $0 \leq n_4 \leq N_b$.    $S^m = N_a -n_4$.

\subsubsection{$  c^{b} \geq 2(N_a+1) c^{ab}$, $c^a \leq 0$}

$S_b^m=0$, $S_a^m=S^m=N_a$.

\subsection{$c^{a}>0$, $c^{b}>0$, $c^{a}+c^{b} < 2 c^{ab} $ \label{apb8} }

The case of $c^{a}>0$, $c^{b}>0$ and $c^{a}c^{b}\leq(c^{ab})^2$ has too many possibilities to be calculated in detail here. We only consider the sub-regime $c^{a}>0$, $c^{b}>0$, $c^{a}+c^{b} < 2 c^{ab} $. According to Equations~(\ref{a15}) and (\ref{a17}), there is always an interval of $S_a$ with  $E=\frac{c^a+c^b-2c^{ab}}{2} S_a(S_a+1)$ and $S_a=S_b$. As $c^{a}+c^{b} < 2 c^{ab} $, it is certain that in this regime, $E$ is minimal at the  upper end of  this interval, although it remained undetermined whether at this point $E$ is the minimal in the whole region of $S_a$ and $S_b$.

Hence it is determined that $S_a^m\neq 0$, $S_b^m \neq 0$, consequently the ground state $|S_a^m, S_b^m,|S_a^m-S_b^m|, |S_a^m-S_b^m|\rangle$ is always entangled.


\begin{thebibliography}{99}
\bibitem{ho1} T.-L. Ho, Phys. Rev. Lett.
81, 742 (1998); T. Ohmi and K. Machida, J. Phys. Soc. Jpn. 67, 1822
(1998).
\bibitem{law} C. K. Law, H. Pu, and N. P. Bigelow, Phys. Rev. Lett.
{\bf 81}, 5257 (1998).
\bibitem{koashi}  M. Koashi and M. Ueda, Phy.
Rev. Lett. {\bf 84}, 1066 (2000).
\bibitem{ho2} T. L. Ho and S. K.
Yip, Phy. Rev. Lett. {\bf 84}, 4031 (2000).
\bibitem{hoyin} T. L. Ho and L. Yin, Phy. Rev. Lett. {\bf 84},
2302 (2000).
\bibitem{yang} K. Yang, arXiv:0907.4739, and references therein.
\bibitem{spinor}
J. Stenger {\it et al.},
Nature {\bf 396}, 345 (1998); H.-J. Miesner {\it et al.}, Phy. Rev.
Lett. {\bf 82}, 2228 (1999); D. M. Stamper-Kurn {\it et al.}, Phy.
Rev. Lett. {\bf 83}, 661 (1999); H. Schmaljohann {\it et al.}, Phy.
Rev. Lett. {\bf 92}, 040402 (2004); M. S. Chang {\it et al.}, Phy.
Rev. Lett. {\bf 92}, 140403 (2004); T. Kuwamoto {\it et al.}, Phy.
Rev. A {\bf 69}, 063604 (2004); M. S. Chang  {\it et al.}, Nature Phys. {\bf 1}, 111 (2005); L. E. Sadler {\it et al.}, Nature {\bf 443}, 312 (2006).
\bibitem{two} See, e.g.,  T. L. Ho and V. B. Shenoy, Phy. Rev.
Lett. {\bf 77}, 3276 (1996); B. D. Esry et al., Phys. Rev. Lett. {\bf 78}, 3594 (1997); H. Pu, and N. P. Bigelow,
Phys. Rev. Lett. {\bf 80}, 1130 (1998);  P. Ao and S. T. Chui,
J. Phys. B {\bf 33}, 535 (2000);  E. Timmermans,
Phys. Rev. Lett. {\bf 81}, 5718 (1998); M. Trippenbach et al.,J. Phys. B {\bf 33}, 4017 (2000).
\bibitem{myatt}  C. J. Myatt, {\it et al.}, Phy. Rev. Lett. {\bf 78}, 586 (1997); D. S. Hall {\it et al.},
Phy. Rev. Lett. {\bf 81}, 1539 (1998); D. S. Hall {\it et al.}, Phy.
Rev. Lett. {\bf 81}, 1543 (1998); G. Modugno  {\it et al.}, Phy. Rev. Lett. {\bf 89},
190404 (2002);  G. Thalhammer {\it et al.}, Phy. Rev.
Lett. {\bf 100}, 210402 (2008);  S. B. Papp and C. E. Wieman, Phy. Rev. Lett. {\bf 97}, 180404 (2006);  S. B. Papp, J. M. Pino and C. E. Wieman, Phy. Rev. Lett. {\bf
101}, 040402 (2008).
\bibitem{shi0} Y. Shi,  {\rm  Int}. J. Mod. Phys. B {\bf 15}, 3007
(2001).
\bibitem{shi1} Y. Shi and Q. Niu, Phy. Rev. Lett. {\bf 96}, 140401
(2006).
\bibitem{shi2} Y. Shi, EPL {\bf 86}, 60008 (2009).
\bibitem{shi4} Y. Shi, Phys. Rev. A {\bf 82}, 013637 (2010).
\bibitem{luo} M. Luo, Z. Li and C. Bao, Phys. Rev. A {\bf 75}, 043609 (2007).
\bibitem{xu1} Z. Xu, Y. Zhang and L. You, Phys. Rev. A {\bf 79}, 023613 (2009).
\bibitem{shi5} Y. Shi, Phys. Rev. A {\bf 82}, 023603(2010).
\bibitem{xu2} The entangled ground state that had been obtained in \cite{shi5}, which had also appeared as a preprint [Y. Shi, e-print arXiv:0912.2209],  only for a specified parameter regime was later  claimed, without a proof, to be the ground state for a large interspecies antiferromagnetic interspecies coupling and irrespective of intraspecies  couplings [Z. Xu {\it et al.}, Phys. Rev. A {\bf 81}, 033603 (2010)], in disagreement with \cite{shi5} and the present paper.
\bibitem{cg} See, for example, A. R. Edmonds, {\em Angular Momentum in Quantum Mechanics}, 2nd edition (Priceton University, Princeton, 1960).
\bibitem{chin} C. Chin, R. Grimm, P. Julienne and E. Tiesinga, Rev. Mod. Phys. {\bf 82}, 1225 (2010).
\end{thebibliography}
\end{document}